\pdfoutput=1

\documentclass[10pt]{paper}

\usepackage{setspace} 
\usepackage{amssymb}
\usepackage{amsfonts}
\usepackage{graphicx}
\usepackage{subfigure}
\usepackage{enumitem}
\usepackage{amsmath}
\usepackage{natbib}
\usepackage{setspace}
\usepackage{fancybox}
\usepackage[bottom]{footmisc}
\usepackage[left=2cm,top=2.5cm,right=3cm,bottom=2cm]{geometry}
\usepackage[normal]{caption}
\usepackage{rotating}
\usepackage{color}
\usepackage{url}
\usepackage[english]{babel}
\usepackage{booktabs}
\usepackage{siunitx}
\usepackage{footnote}
\usepackage{footmisc}
\usepackage{multirow}
\usepackage{fancyhdr}
\usepackage{verbatim}
\usepackage[usenames,dvipsnames]{xcolor}
\usepackage{pifont}
\usepackage{relsize}
\definecolor{lightgreen}{rgb}{0,1,0}
\definecolor{darkgray}{gray}{0.20}
\usepackage{longtable}
\usepackage{changes}
\usepackage{xcolor}
\usepackage{lineno}

\makeatletter
\renewcommand{\@biblabel}[1]{\quad#1.}
\makeatother

\date{}
\begin{document}


\begin{flushleft}
{\Large
\textbf{Organization and hierarchy of the human functional brain network lead to a chain-like core}
}

Rossana Mastrandrea$^{*1}$, Andrea Gabrielli$^{1,2}$, Fabrizio Piras$^{3,4}$, Gianfranco Spalletta$^{4,5}$, Guido Caldarelli$^{1,2}$ Tommaso Gili$^{3,4}$
\\
\bf{1} IMT School for Advanced Studies, Lucca, piazza S. Ponziano 6, 55100 Lucca, Italy
\\
\bf{2} Istituto dei Sistemi Complessi (ISC) - CNR, UoS Sapienza, Dipartimento di Fisica, Universit\'a \lq\lq Sapienza\rq\rq; P.le Aldo Moro 5, 00185 - Rome, Italy
\\
\bf{3} Enrico Fermi Center, Piazza del Viminale 1, 00184 Rome, Italy
\\
\bf{4} IRCCS Fondazione Santa Lucia, Via Ardeatina 305, 00179 Rome, Italy
\\
\bf{5} Menninger Department of Psychiatry and Behavioral Sciences, Baylor College of Medicine, Houston, Tx, USA
\\

$*$ E-mail: rossana.mastrandrea@imtlucca.it
\\
\end{flushleft}

\section*{Abstract} 
The brain is a paradigmatic example of a complex system as its functionality emerges  as a global property of local mesoscopic and microscopic interactions. Complex network theory allows to elicit the functional architecture of the brain in terms of links (correlations) between nodes (grey matter regions) and to extract information out of the noise. Here we present the analysis of functional magnetic resonance imaging data from forty  healthy humans during the resting condition for the investigation of the basal scaffold of the functional brain network organization.  We show how brain regions tend to coordinate by forming a highly hierarchical chain-like structure of homogeneously clustered anatomical areas. A maximum spanning tree approach revealed the centrality of the occipital cortex and the peculiar aggregation of cerebellar regions to form a closed core. We also report the hierarchy of network segregation and the level of clusters integration as a function of the connectivity strength between brain regions. 
\newline
\newline

\section*{Introduction} 

The intrinsic functional architecture of the brain and its changes due to cognitive engagement, ageing and diseases are nodal topics in neuroscience, attracting considerable attention from many disciplines of scientific investigation. Complex network theory \citep{caldarelli2007scale, barabasi2009scale,newman2010networks} provides tools representing the state-of-the-art of multivariate analysis of local cortical and subcortical mesoscopic interactions. 
The new amount of data available from a variety of different sources showed clearly that the complex network description of both the structural and the functional organization of the brain demonstrates a repertoire of unexpected properties of brain connectomics \citep{bullmore2009complex}. Accordingly network theory allows a description of brain architecture in terms of patterns of communication between brain regions, treated as evolving networks and associate this evolution to behavioral outcomes \citep{bassett2011dynamic}. Brain networks are characterized by a balance of dense relationships among areas highly engaged in processing roles, as well as sparser relationships between regions belonging to systems with different processing roles. This segregation facilitates communication among brain areas that may be distributed anatomically but are needed for sets of processing operations \citep{sporns2013network}. Along with that the integrated functional organization of the brain involves each network component executing a discrete cognitive function mostly autonomously or with the support of other components where the computational load in one is not heavily influenced by processing in the others \citep{bertolero2015modular}.
In this quest for a constantly improving quantitative description of the brain and of the cardinal features of its functioning complex networks plays a crucial role \citep{rubinov2010complex,deco2011emerging,van2011rich,sporns2012simple}. Specifically, it proved to be able to elicit both the scaffold of the mutual interactions among different areas in healthy brains \citep{bullmore2009complex,bassett2016small} and the local failure of the global functioning in diseased brains \citep{Bassett2009,rosazza2011resting,aerts2016brain}.
Network representation describes the brain as a graph with a set of nodes - a variable number of brain areas (from $10\sp{2}$ to $10\sp{4}$) -  connected with links representing functional, structural or effective interactions. The use of complex network theory passed progressively from an initial assessment of basic topological properties
\citep{salvador2005neurophysiological,van2008small,telesford2010reproducibility}  to a more sophisticated description of global features of the brain, as small-worldness \citep{achard2006resilient}, rich-club organization \citep{van2011rich} and topology \citep{petri2014homological,tiz2016}.
 However, a clear understanding of the functional advantage of a network organization in the brain, the characterization of its substrate and a description of the network structure as a function of the level of brain regions interaction are still missing.  
In this paper, we investigate resting state functional networks, where links represent the strength of correlation between time series of spontaneous neural activity as measured by blood oxygen-level-dependent (BOLD) functional MRI (fMRI) \citep{eguiluz2005}. Specifically we interpret a functional connection between two nodes as the magnitude of the synchrony of their low-frequency oscillations, which is associated with the modulation of activity in one node by another one \citep{wang2012electrophysiological,honey2012slow} largely constrained by anatomical connectivity \citep{hagmann2008mapping,honey2009predicting}. 
A percolation analysis of the functional network \citep{gallos2012small,tiz2016} is used to highlight the progressive engagement of brain regions in the whole network as a function of the connectivity strength. Subsequently,
by means of the maximum spanning forest (MSF) and the maximum spanning tree (MST) representations \citep{caldarelli2007scale,caldarelli2012network} we obtain the basal scaffold of the brain network, that shows to be characterized by a linear backbone in which few nodes (cerebellar and occipital regions) play a central role, even progressively increasing the spatial resolution or reducing nodes size.

\section*{Results}

\subsection*{Percolation Analysis}

\begin{figure}[!ht]
\centering
\subfigure[]
{\includegraphics[width=0.7\textwidth]{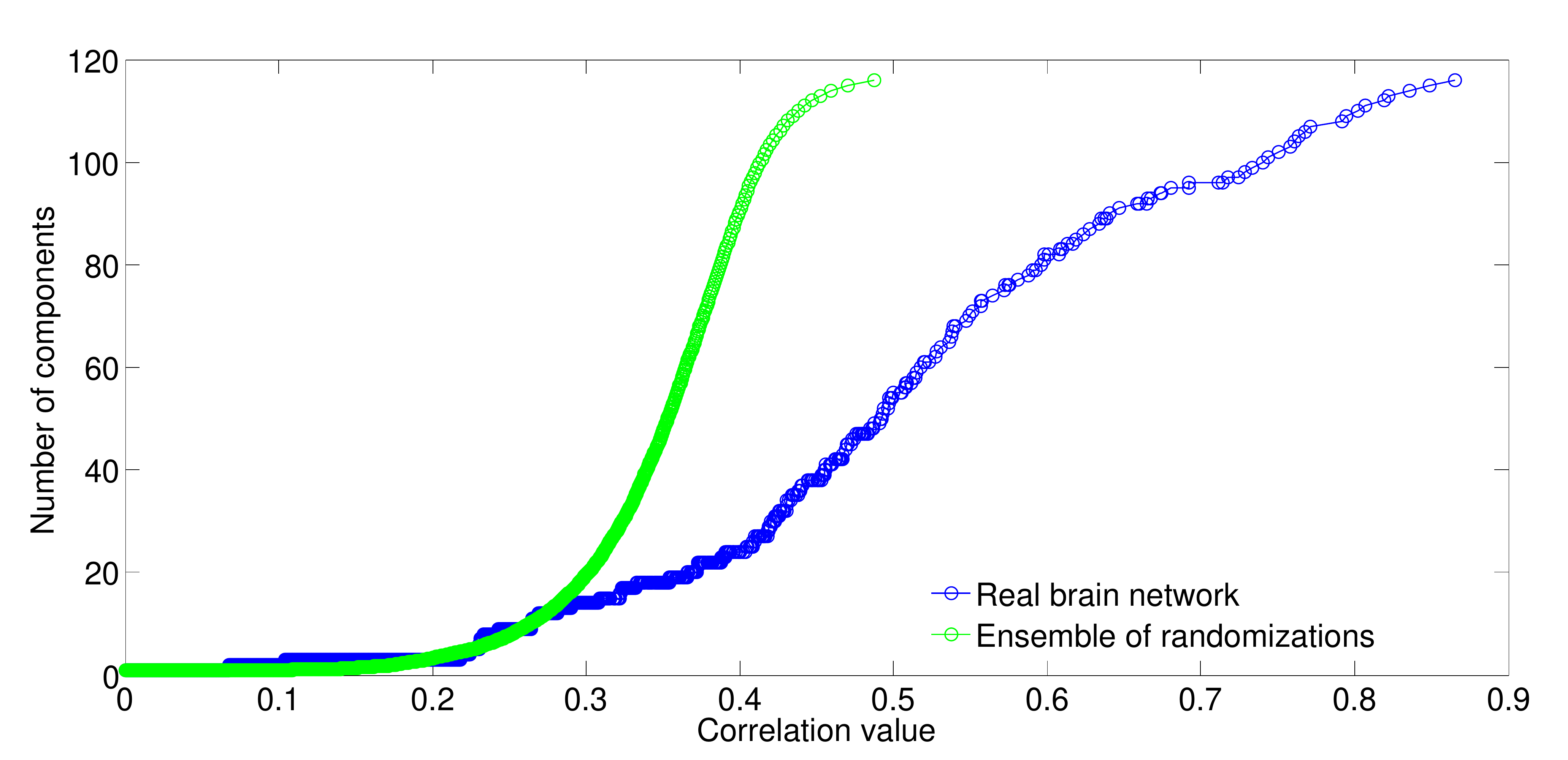}}
\hspace{1mm}
\hspace{-5mm}
\subfigure[ ]
{\includegraphics[width=0.705\textwidth]{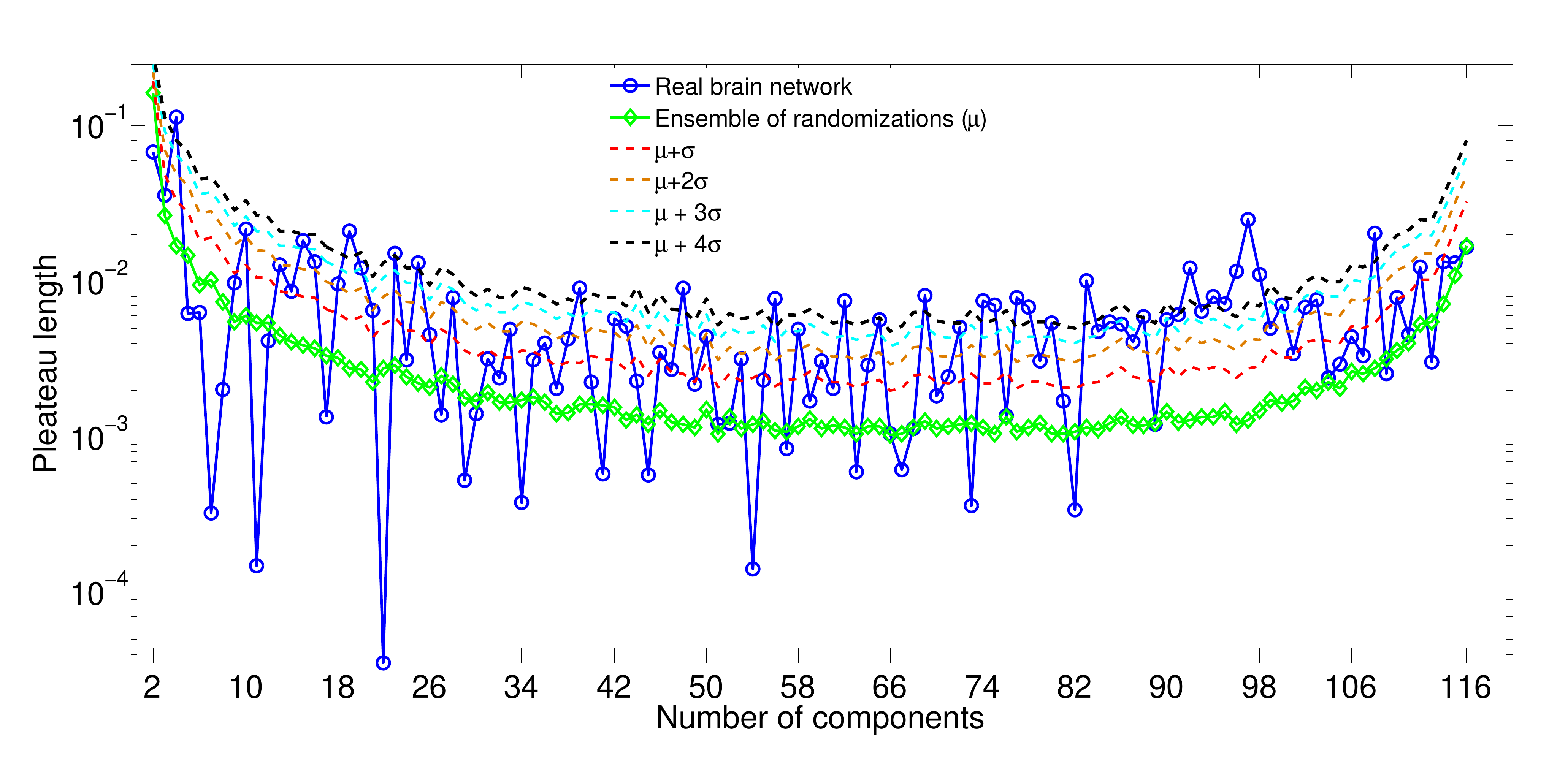}}
\caption{{\bf Percolation analysis.} (a) Distribution of the number of network components versus the threshold applied to the correlation matrix. (b) Distribution of plateaux length observed in figure 1(a) and computed as the increment of correlation value between two newborn components. All figures show the percolation curves for the real case (blue) and the ensemble of its randomizations (green). Moreover figure (b) reports upper confidence intervals: mean plus  respectively, one (red), two (orange), three (light blue) and four (bleak) standard deviations.
\label{perc}}
\end{figure}

The representative  human functional brain network is often analyzed introducing specific thresholds to map the fully-connected correlation matrix in a sparse binary matrix \citep{bullmore2009complex}. Here, to avoid any arbitrary assumption we perform a percolation analysis \citep{gallos2012small,tiz2016} on the whole network. We rank correlations in increasing order,  one at a time we remove from the network the link corresponding to the observed value and explore the  global organization of the remaining network. Specifically,  in fig.\ref{perc} (a) we show the number of connected components updated each time a new link is removed. 
The emergence and the number of \textit{plateaux} shed light on the hierarchical structure of the network, their length unveils the intrinsic stability of certain network configurations after the removal of links.
For the human functional brain network a remarkable hierarchy in the disaggregation process emerges from the comparison with a proper null model. The same percolation analysis is performed on the ensemble of 100 randomizations (Methods) of the observed correlation matrix showing a  faster disaggregation in disconnected components with plateaux absent or of small length.
\


We compute the distribution of plateaux length looking at the increment of correlation values when the network passes from $n$ to $n+1$ components  and show it in fig.\ref{perc} (b). In the same figure we also report the  average plateaux length computed on the ensemble of randomizations.  A great variability characterizes the percolation curve of the real network with significant deviations of the plateaux length from the random case. 

\newpage

\section*{Chain-like modules}

\begin{figure}[!ht]
\centering
{\includegraphics[width=1\textwidth]{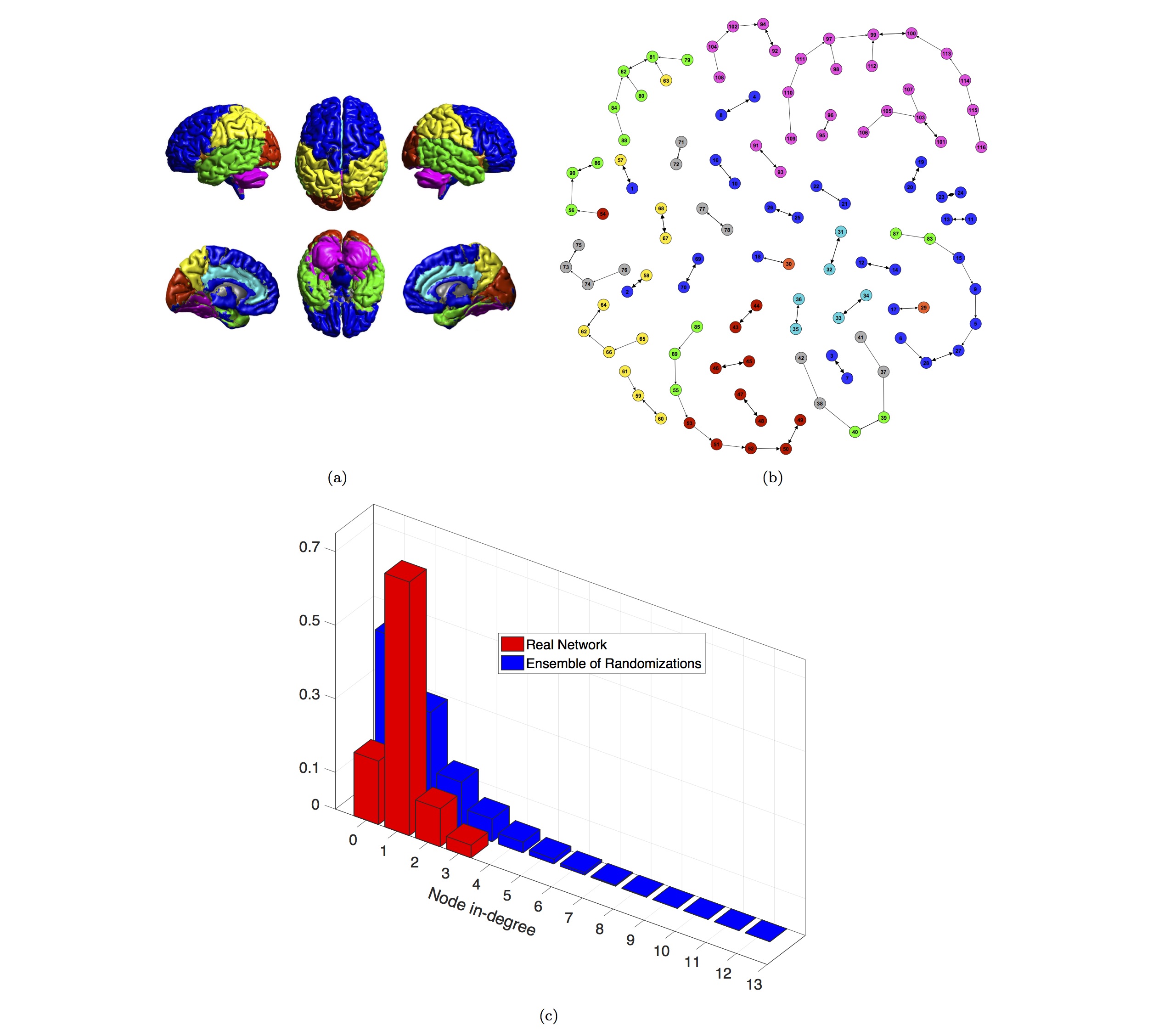}}
\caption{{\bf Maximum Spanning Forest of the real human functional brain network.} (a) Anatomical grouping of AAL parcellation of the human brain.
 $\textcolor{Blue}{\mathlarger{\mathlarger{\mathlarger{\bullet}}}}$  Frontal Lobe; $\textcolor{Orange}{\mathlarger{\mathlarger{\mathlarger{\bullet}}}}$  Insula; 
$\textcolor{Cyan}{\mathlarger{\mathlarger{\mathlarger{\bullet}}}}$ Cingulate; 
$\textcolor{LimeGreen}{\mathlarger{\mathlarger{\mathlarger{\bullet}}}}$ Temporal Lobe; $\textcolor{Red}{\mathlarger{\mathlarger{\mathlarger{\bullet}}}}$  Occipital Lobe; $\textcolor{Yellow}{\mathlarger{\mathlarger{\mathlarger{\bullet}}}}$ Parietal Lobe;  $\textcolor{Gray}{\mathlarger{\mathlarger{\mathlarger{\bullet}}}}$ Deep Grey Matter;  $\textcolor{VioletRed}{\mathlarger{\mathlarger{\mathlarger{\bullet}}}}$ Cerebellum. (b) Maximum spanning forest components. Arrows indicate that the ROI-source is maximally correlated with the ROI-target. (c) Node in-degree distribution of the MSF. Comparison between the real brain network and the ensemble of its randomizations (100).
\label{MSF}}
\end{figure}

The percolation analysis highlights the existence of a not trivial functional organization of the human brain network. Here, we investigate  it considering  a filtered network where for each area all weighted links but the strongest are discarded. This approach  gives rise to 36 disconnected components  forming a Maximum Spanning Forest (Methods) with a new information on link directionality. It simply indicates that each source points toward its  maximally correlated brain area and it is not necessarily reciprocated.  Figure \ref{MSF} shows the abundance of modules with size 2 and reciprocated links: those are meanly mirror-areas of the right and left brain hemispheres or adjacent/very close ROIs belonging to the same hemisphere. Nodes are coulored according to the anatomical regions reported in figure \ref{MSF} (a) (detailed names of ROIs in table 1 in SI). A noteworthy result concerns groups of size greater than 3 exhibiting a chain-like structure, sometimes very long as for the Cerebellum.  This implies that most of the nodes in the MSF have in-degree\footnote{Out-degree is equal to 1 for construction.} equal to one, few equal to 2 and very few greater than 2. Furthermore,  nodes tend to connect with nodes belonging to the same anatomical region. The only exception is represented by the Temporal Lobe: ROIs in this region are linked with all the other anatomical areas except for the Cerebellum and the Deep Grey Matter ones.

We build the MSF of the 100 randomizations of the real network obtaining a number of components varying in the range $[12,24]$. All randomized correlation matrices exhibit a star-like organization of components in the MSF, while the number of modules of size 2 is dramatically reduced. Moreover, colors of linked nodes are completely random. In figure S1 we show the MSF performed on one of such randomizations. We compute the distribution of node in-degree of the observed MSF and of the ensemble of its randomizations (fig. 2(c)). In the real case the in-degree is small with almost the $70\%$ of  values equal to 1. The ensemble of randomizations shows more variability and a left-skewed distribution of in-degree, with almost the $50\%$ of values equal to 0, the $30\%$ equal to 1 and a maximum of 15. The outcome confirms that the MSF of all the randomized correlation matrices shows similar features: star-like organization of modules with few in-degree hubs representing the strongest connected partner for several brain areas. This result highlights the fundamental hierarchical organization of brain functional areas.

\newpage
\subsubsection*{From forest to tree}

As next step, we link together all groups of nodes in fig.\ref{MSF}(b) such that (i) we add only one link between two modules previously disconnected and with the greatest weight; (ii) we end up with a unique connected component (Methods). The resulting network is a Maximum Spanning Tree as all nodes are connected through a  maximal weighted path without forming loops.

\begin{figure}[!ht]
\centering
{\includegraphics[width=0.8\textwidth]{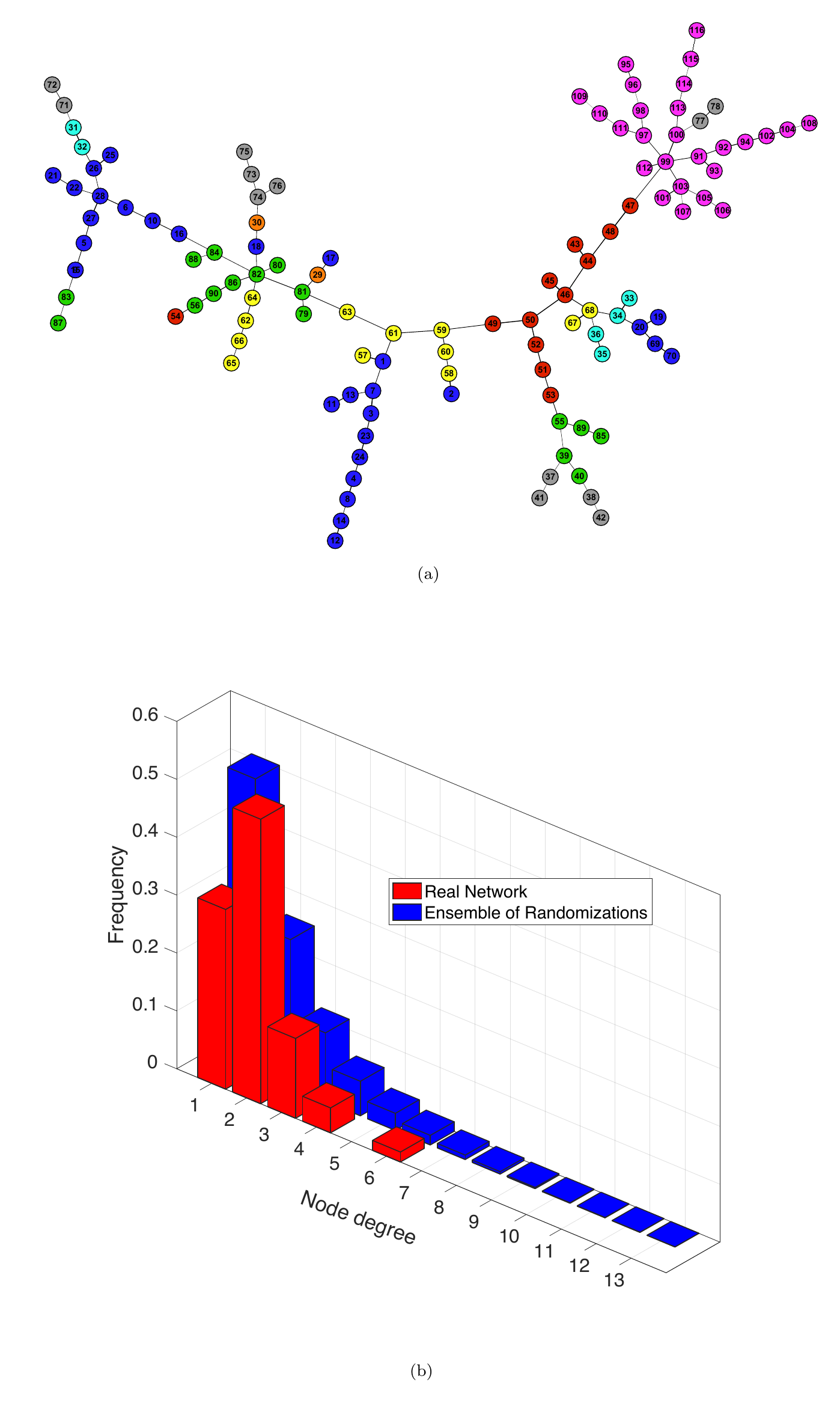}}
\caption{{\bf Maximum Spanning Tree of the real human functional brain network.} (a) Colors represent anatomical regions according to the grouping of AAL parcellation in fig.\ref{MSF}(a) :$\textcolor{Blue}{\mathlarger{\mathlarger{\mathlarger{\bullet}}}}$  Frontal Lobe; $\textcolor{Orange}{\mathlarger{\mathlarger{\mathlarger{\bullet}}}}$  Insula; 
$\textcolor{Cyan}{\mathlarger{\mathlarger{\mathlarger{\bullet}}}}$ Cingulate; 
$\textcolor{LimeGreen}{\mathlarger{\mathlarger{\mathlarger{\bullet}}}}$ Temporal Lobe; $\textcolor{Red}{\mathlarger{\mathlarger{\mathlarger{\bullet}}}}$  Occipital Lobe; $\textcolor{Yellow}{\mathlarger{\mathlarger{\mathlarger{\bullet}}}}$ Parietal Lobe;  $\textcolor{Gray}{\mathlarger{\mathlarger{\mathlarger{\bullet}}}}$ Deep Grey Matter;  $\textcolor{VioletRed}{\mathlarger{\mathlarger{\mathlarger{\bullet}}}}$ Cerebellum. (b) Node degree distribution in the MST. Comparison between the real brain network and the ensemble of its randomizations (100).
\label{Tree}}
\end{figure}

The MST in fig.\ref{Tree}(a) preserves the chain-like organization of nodes and mainly reproduces the anatomical division in regions. Some star-like aggregations emerge revealing the centrality of certain nodes as ROI 99, belonging to the Cerebellum and 82, part of the Temporal Lobe. Remarkable is the separation of the Deep Grey Matter, which represents the ancient part of the human brain, connecting to the periphery of the tree with the Temporal  Lobe, the Insula and Cingulate areas. Nor chain-like organization neither relevant agglomeration of nodes are observable in the tree obtained from the MSF of the randomized correlation matrices (one examples in fig. S2). Also in this case the distribution of node degree characterizes the two kinds of network organization: star-like (null model) and chain-like (real brain network). The comparison between the real network and the ensemble of its randomization (fig. 3(b)) confirms the star-like organization of the random MST, with half of nodes having degree equal to 1 and maximum degree equal to 17.

\subsection*{Hierarchical integration of brain areas  }

The maximum spanning tree approach 
doesn't provide any evidence about the internal connectivity  of components observed in the MSF, as loops are not allowed. Here we showed a number of snapshots of the network associated with specific percolation phases. By selecting appropriate thresholds and discarding correlation values below them, we inspected the network formation process according to the strength of edges. The rationale behind this approach comes from the possibility to explore the level of segregation of brain areas and how they functionally and hierarchically  integrate. Thresholds are chosen looking at figure \ref{perc} (b) and considering only correlation values associated to significant deviations of plateaux length of the real network from the random case.

\begin{figure}[!ht]
\centering
\subfigure[Threshold = 0.77]
{\includegraphics[width=0.5\textwidth]{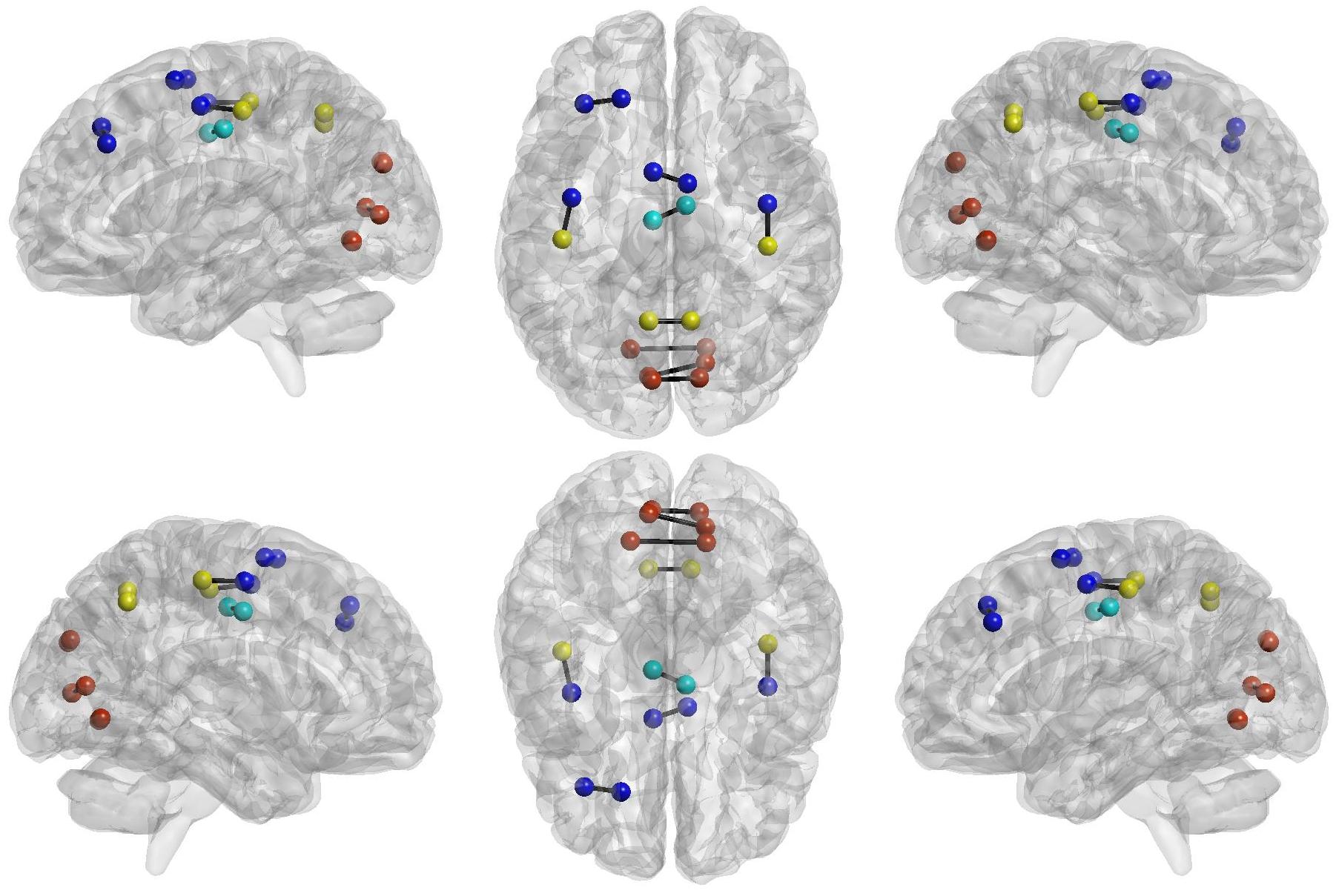}}
\vspace{7mm}
\hspace{8mm}
{\includegraphics[width=0.28\textwidth]{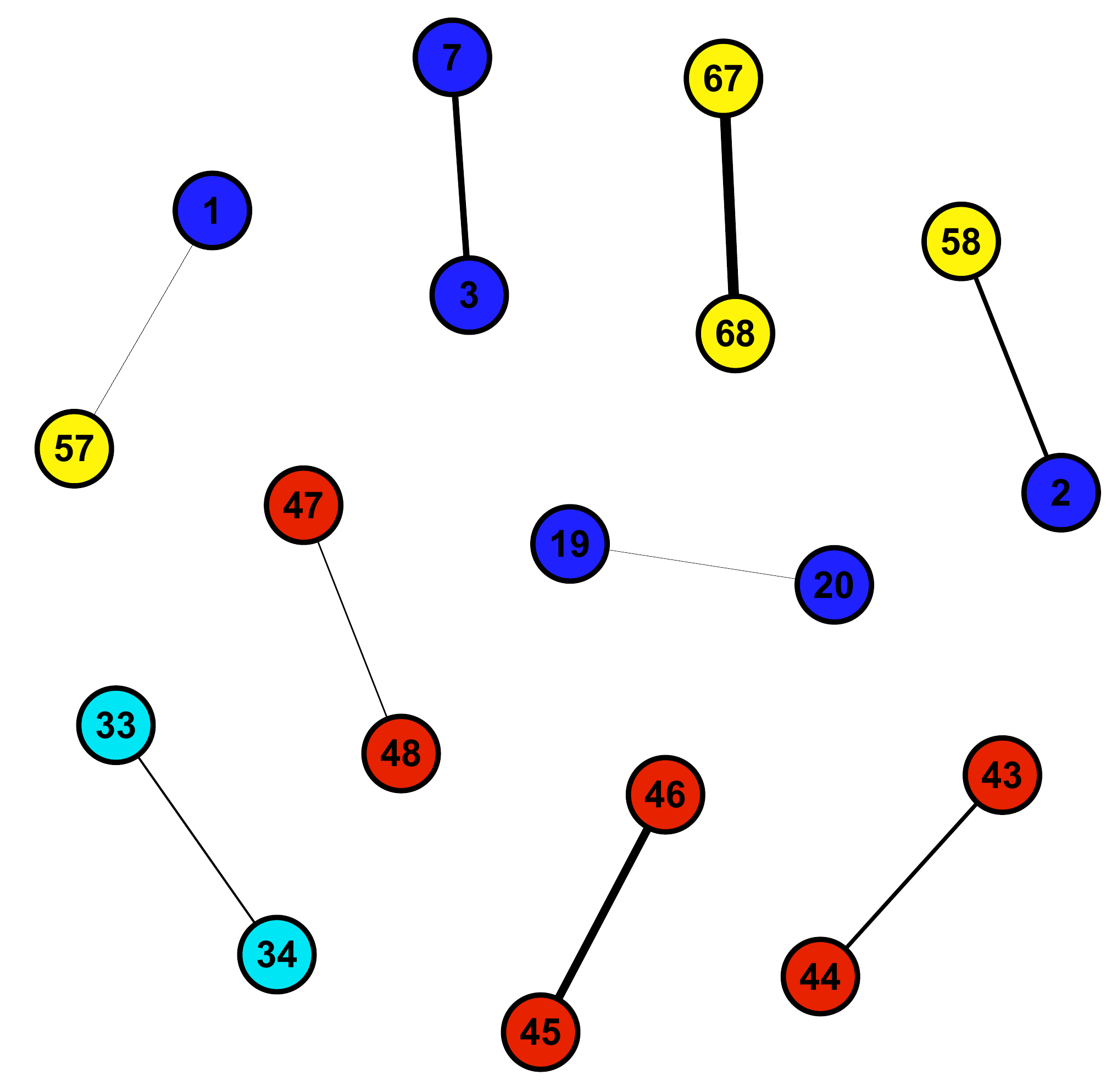}}
\subfigure[Threshold = 0.72]
{\includegraphics[width=0.5\textwidth]{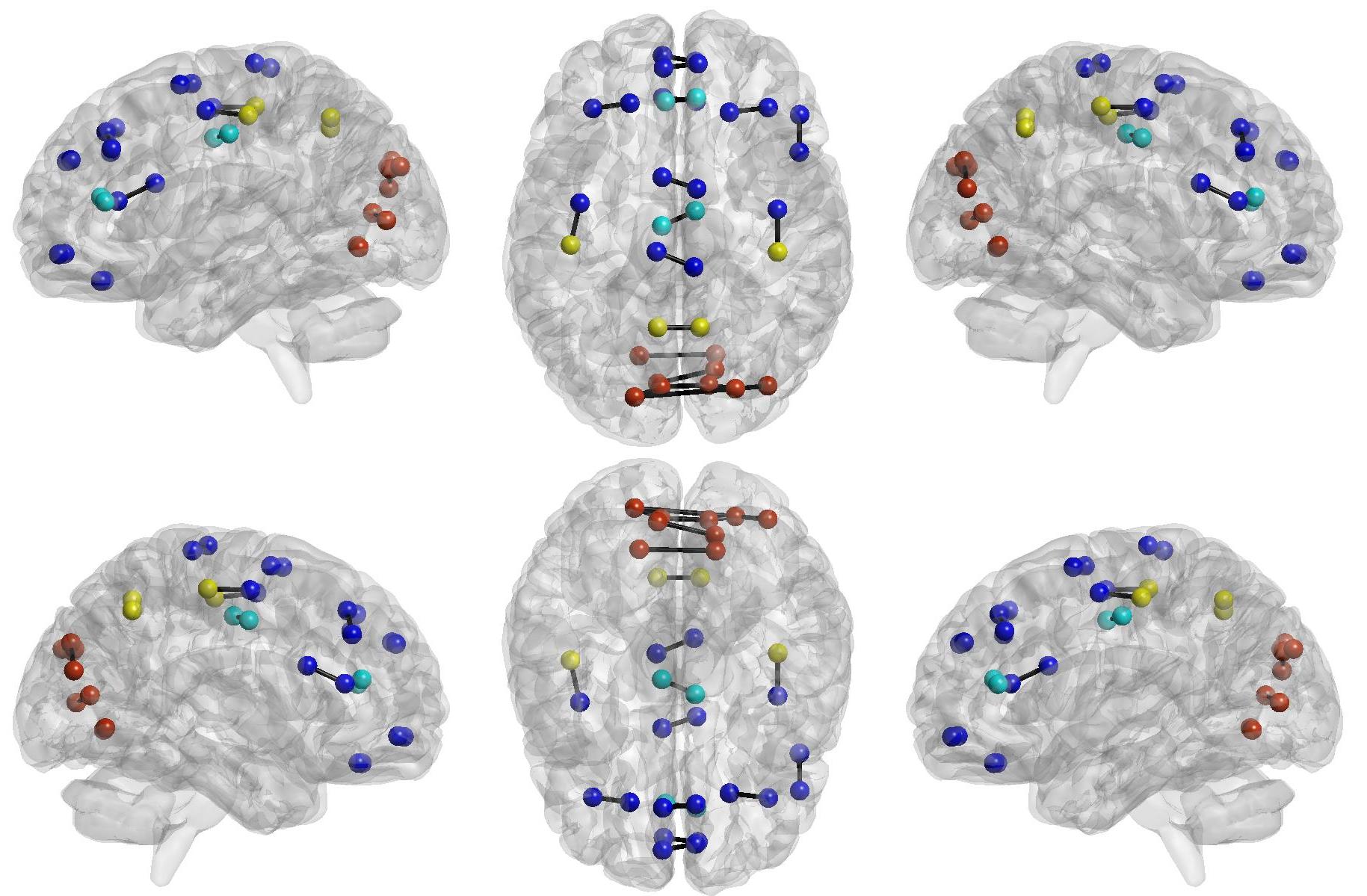}}
\vspace{7mm}
\hspace{8mm}
{\includegraphics[width=0.3\textwidth]{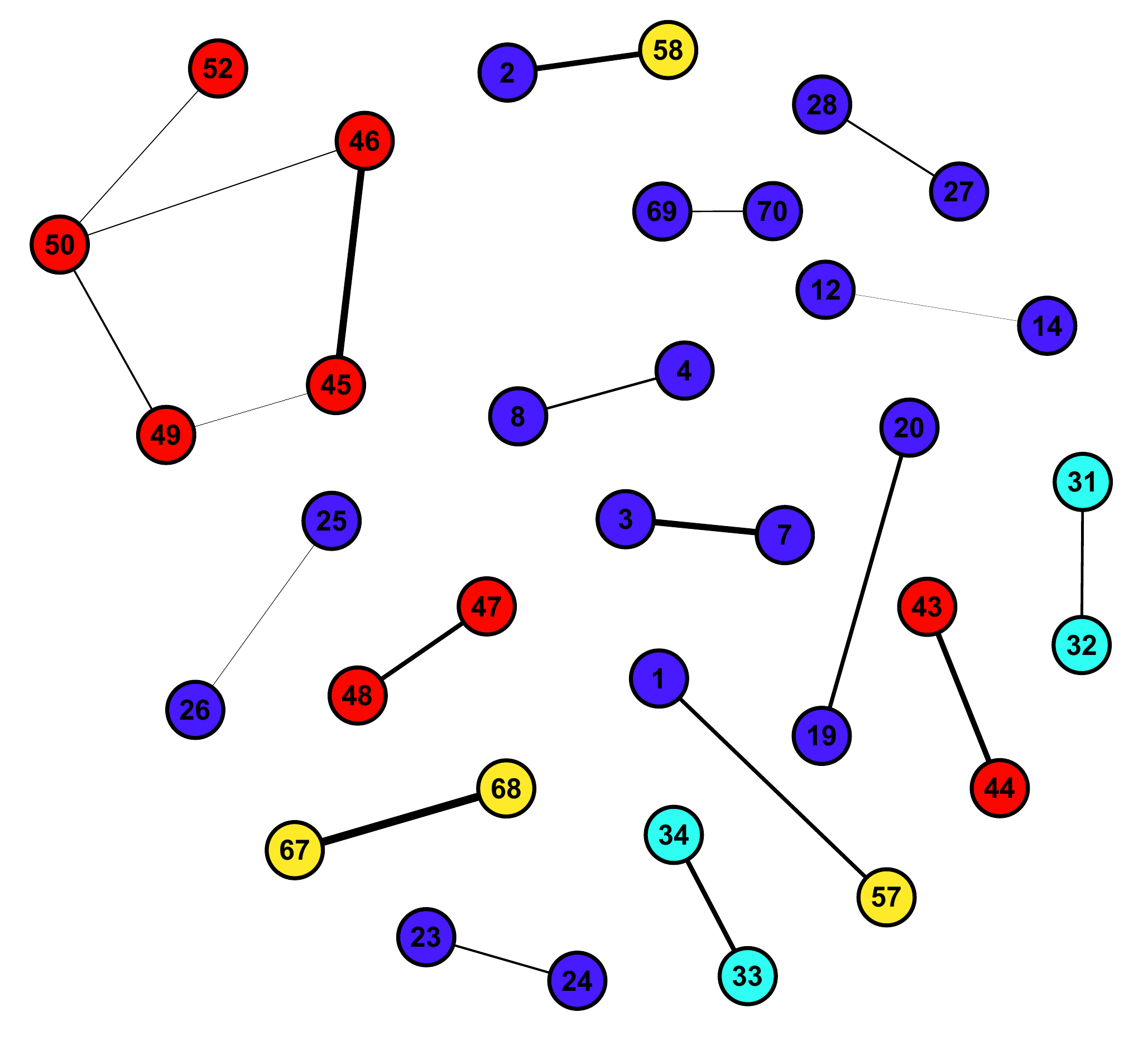}}
\hspace{3mm}
\subfigure[Threshold = 0.68 ]
{\includegraphics[width=0.5\textwidth]{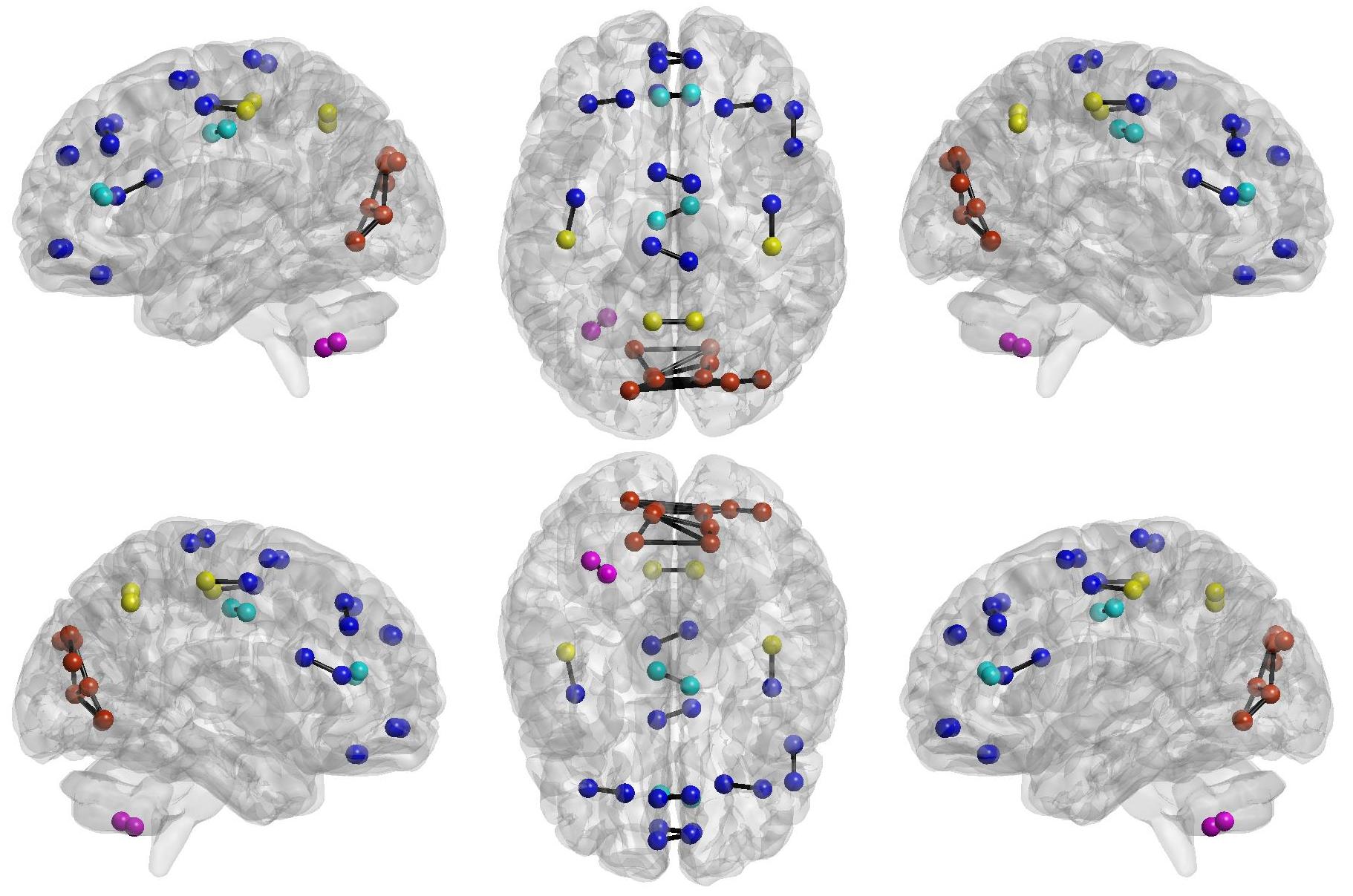}}
\hspace{8mm}
{\includegraphics[width=0.3\textwidth]{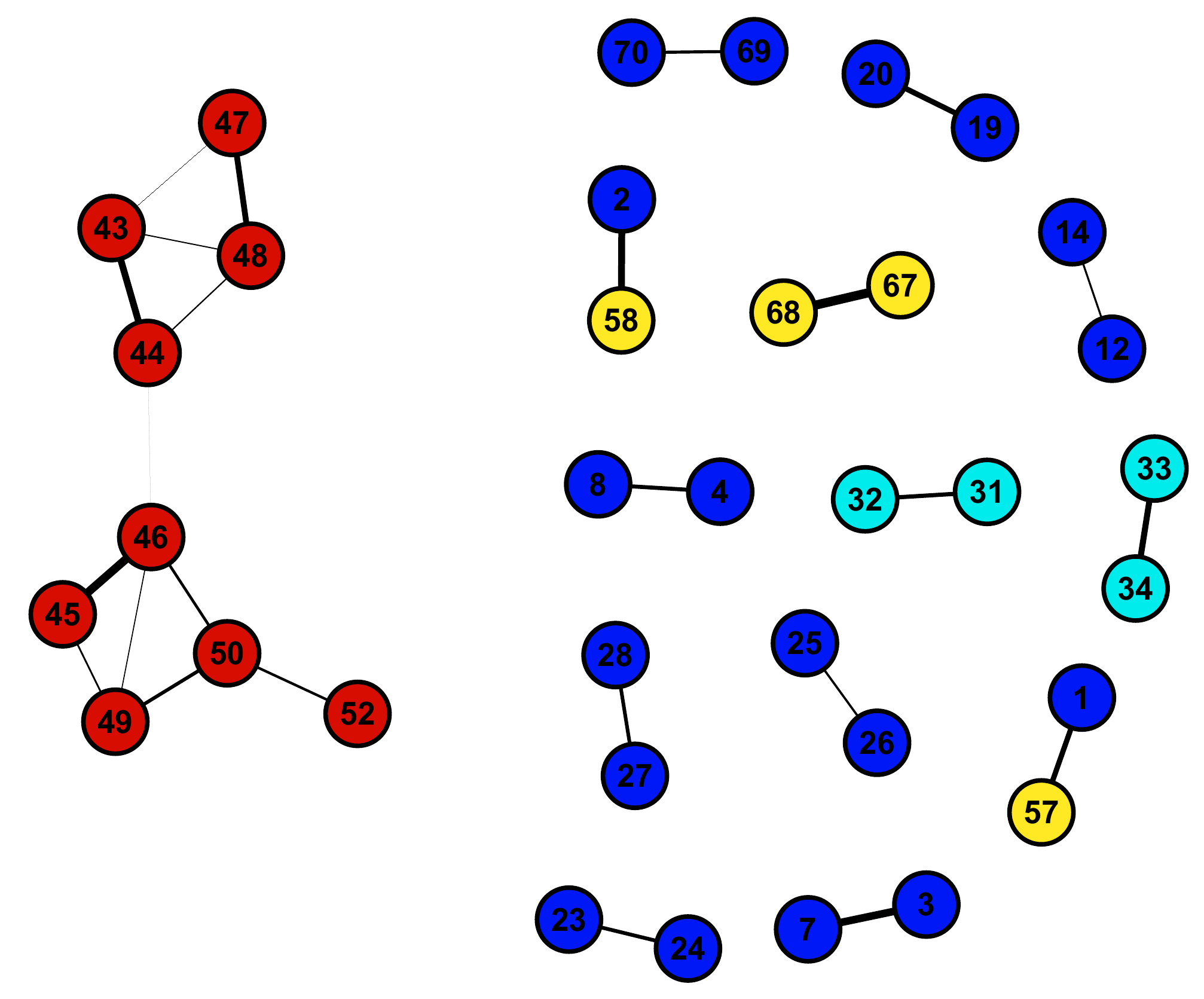}}
\caption{{\bf Snapshots of the human functional brain network.} Each snapshot is realized thresholding the correlation matrix. Thresholds are chosen looking at the most significant deviation of plateaux length of the real percolation curve from its ensemble of randomization (fig.\ref{perc}(b). ) Colors represent anatomical regions according to the grouping of AAL parcellation in fig.\ref{MSF}(a): $\textcolor{Blue}{\mathlarger{\mathlarger{\mathlarger{\bullet}}}}$  Frontal Lobe; $\textcolor{Orange}{\mathlarger{\mathlarger{\mathlarger{\bullet}}}}$  Insula; 
$\textcolor{Cyan}{\mathlarger{\mathlarger{\mathlarger{\bullet}}}}$ Cingulate; 
$\textcolor{LimeGreen}{\mathlarger{\mathlarger{\mathlarger{\bullet}}}}$ Temporal Lobe; $\textcolor{Red}{\mathlarger{\mathlarger{\mathlarger{\bullet}}}}$  Occipital Lobe; $\textcolor{Yellow}{\mathlarger{\mathlarger{\mathlarger{\bullet}}}}$ Parietal Lobe;  $\textcolor{Gray}{\mathlarger{\mathlarger{\mathlarger{\bullet}}}}$ Deep Grey Matter;  $\textcolor{VioletRed}{\mathlarger{\mathlarger{\mathlarger{\bullet}}}}$ Cerebellum.
\label{MSTsnap}}
\end{figure}

Panels from (a) to (n) in fig.\ref{MSTsnap} show the human functional brain network related to the aforementioned thresholds. As the threshold value decreases (from (a) to (n)), the number of links increases showing how groups of nodes appear and their between and within connections intensify.  The sequence of snapshots reveal the hierarchical functional organization of the  brain network. Nodes do not connect forming several large and disconnected components, but: (i)  nodes belonging to the Occipital Lobe start to connect each other and the anatomical region is almost completely connected after few steps;  (ii) some nodes belonging to the same region form small chains; (iii) several nodes directly  connect  to the Occipital Lobe.  Therefore, the most central area is represented by the Occipital Lobe, while ROIs belonging to the Deep Grey Matter lie in the periphery connecting to the network at the last moment in the percolation phases. The Cerebellum forms a quite long chain before joining the Occipital Lobe through the ROI 99 already emerged in the MST for its centrality.

\begin{figure}[!ht]
\subfigure[Threshold = 0.58]
{\includegraphics[width=0.5\textwidth]{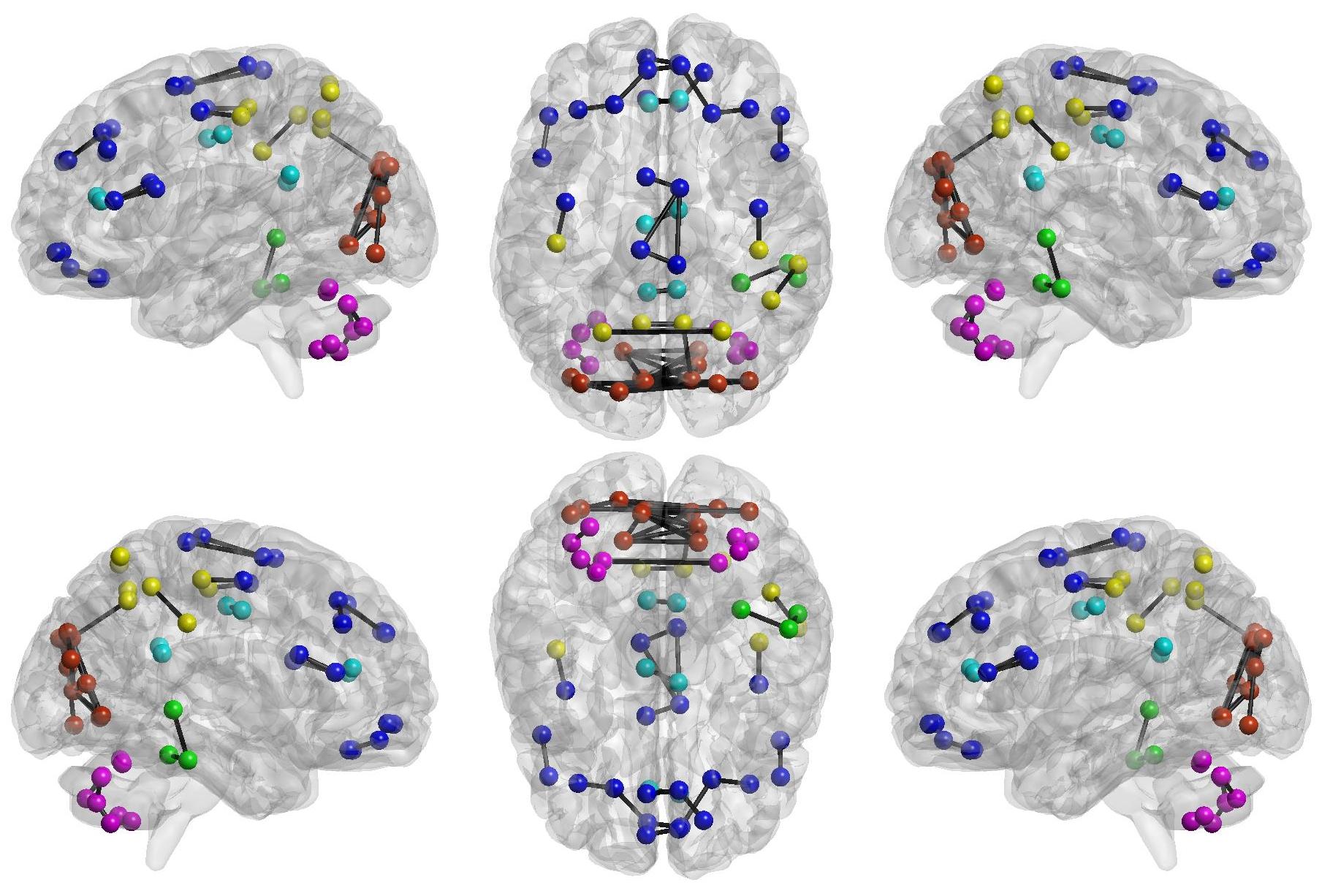}}
\hspace{1mm}
{\includegraphics[width=0.5\textwidth]{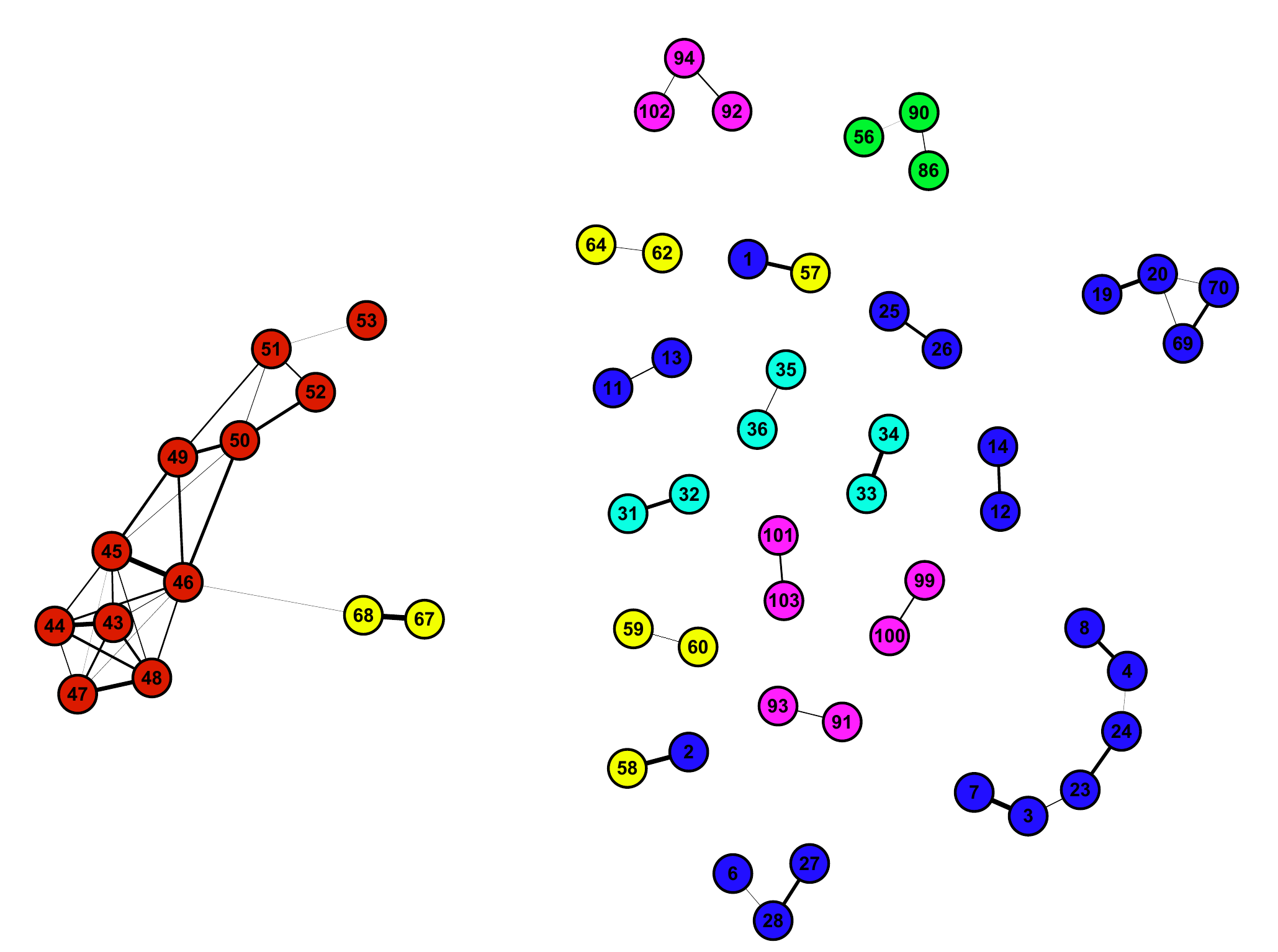}}
\subfigure[Threshold = 0.54]
{\includegraphics[width=0.5\textwidth]{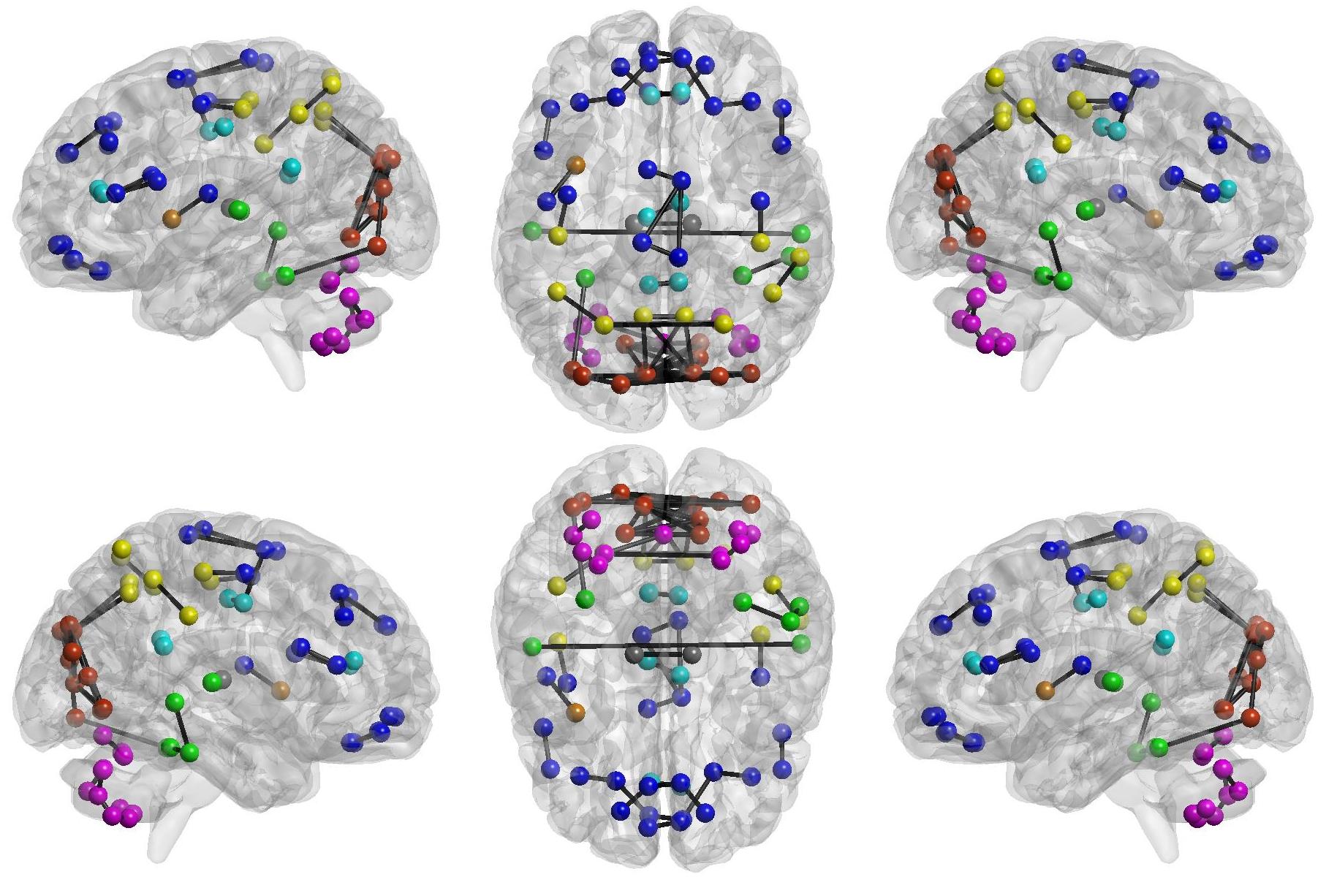}}
\hspace{1mm}
{\includegraphics[width=0.52\textwidth]{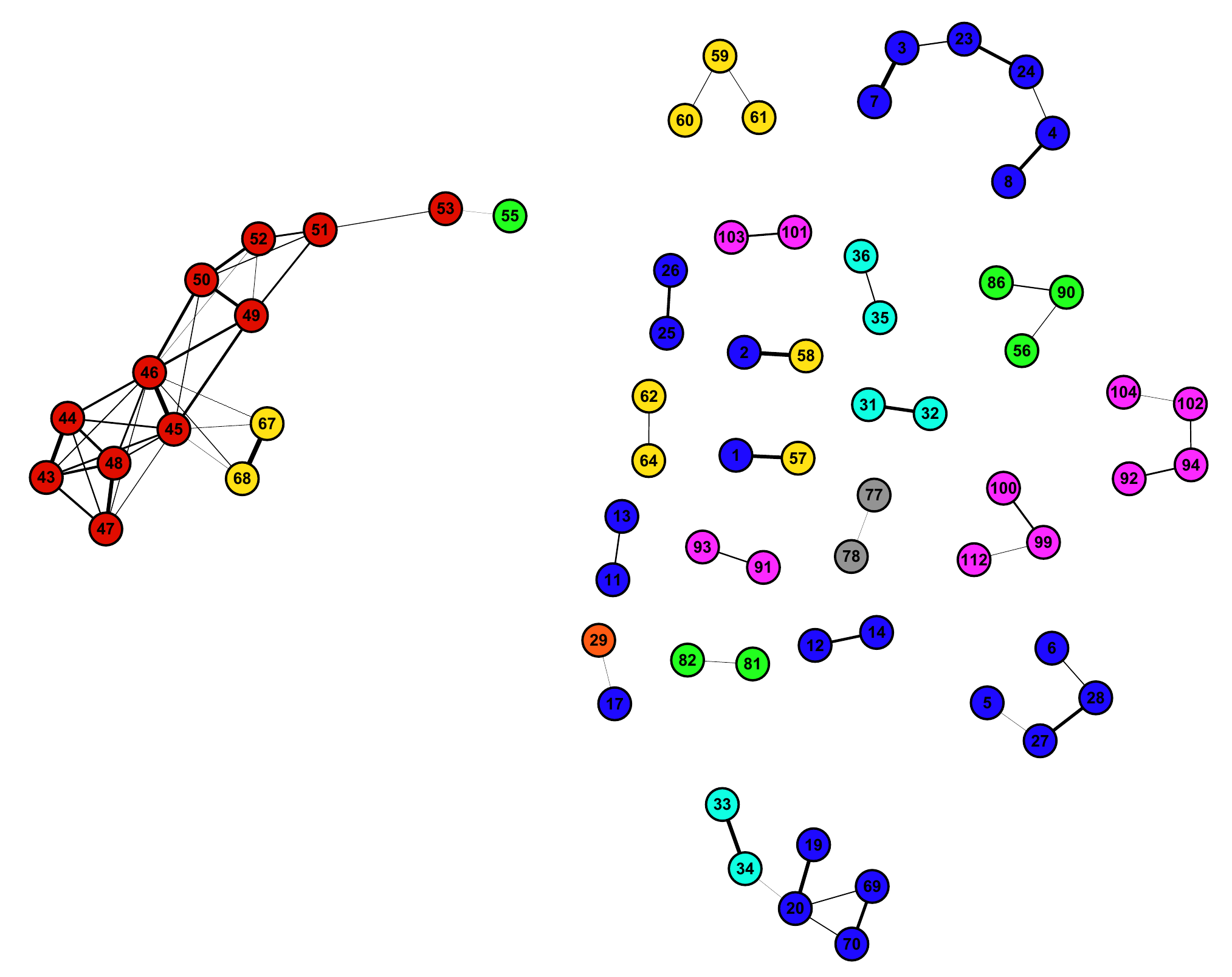}}
\subfigure[Threshold = 0.52]
{\includegraphics[width=0.5\textwidth]{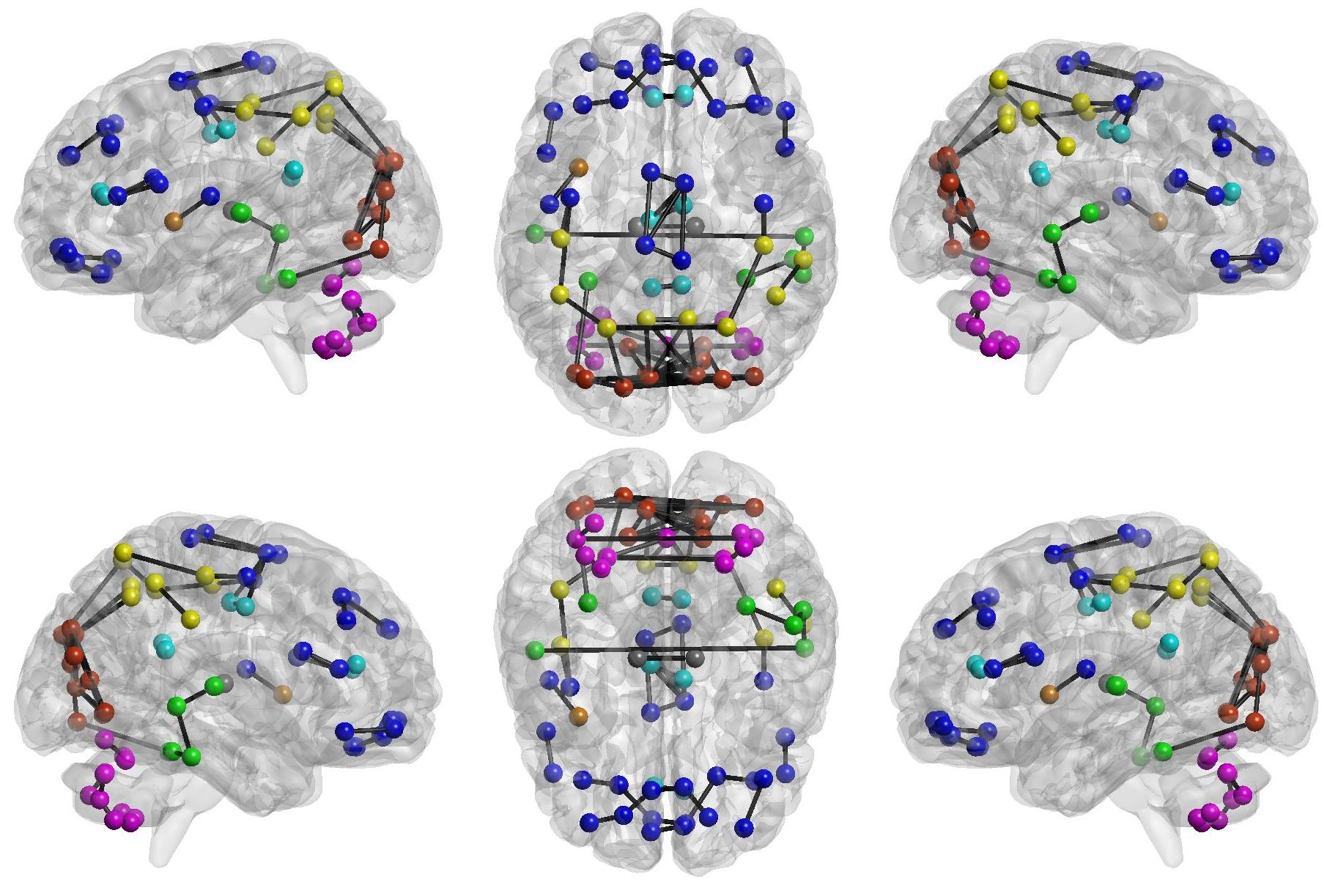}}
\hspace{5mm}
{\includegraphics[width=0.4\textwidth]{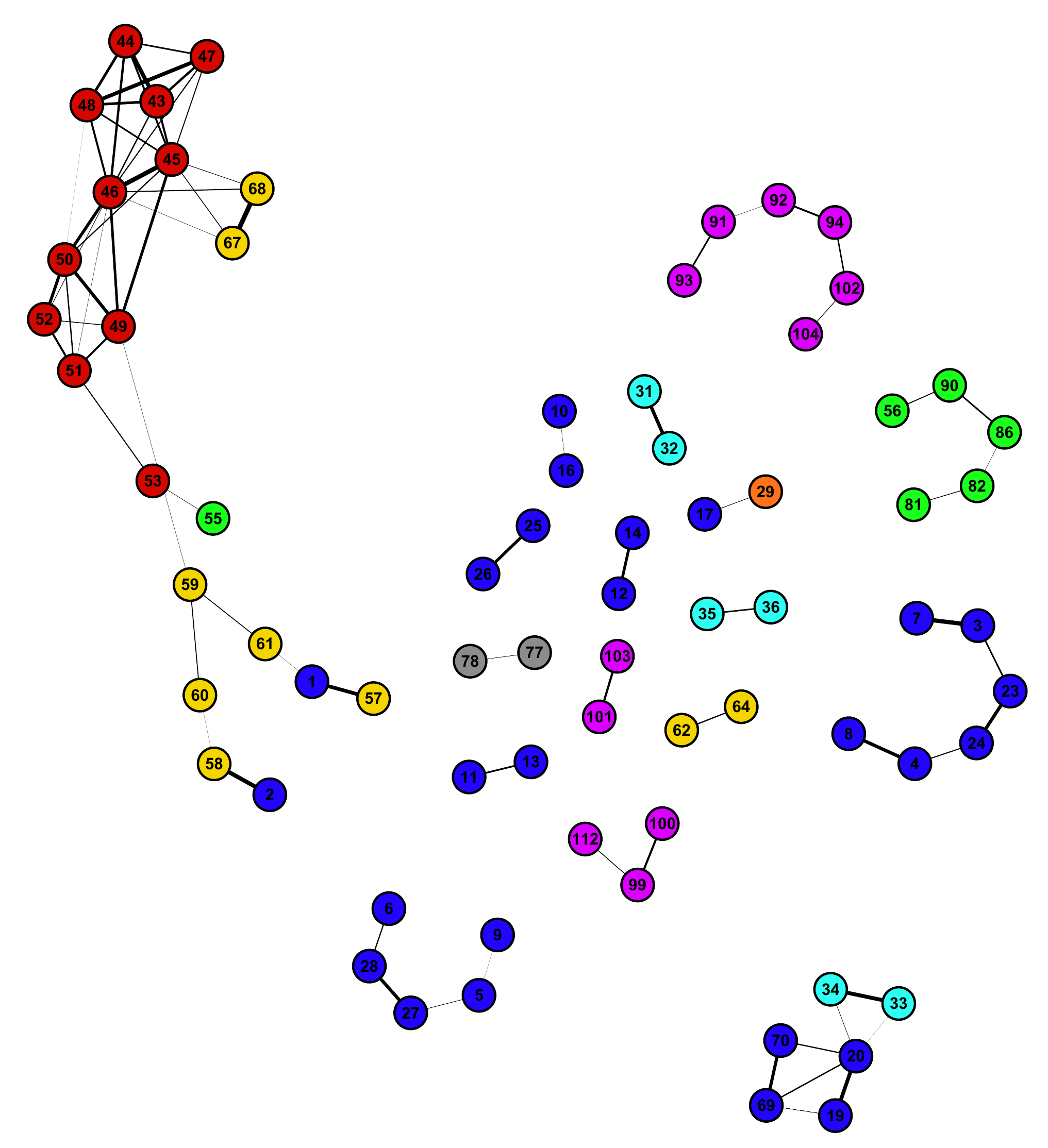}}
\caption{{\bf Snapshots of the human functional brain network.} Each snapshot is realized thresholding the correlation matrix. Thresholds are chosen looking at the most significant deviations of plateaux length of the real percolation curve from its ensemble of randomizations (fig.\ref{perc}(b).) Colors represent anatomical regions according to the grouping of AAL parcellation in fig.\ref{MSF}(a): $\textcolor{Blue}{\mathlarger{\mathlarger{\mathlarger{\bullet}}}}$  Frontal Lobe; $\textcolor{Orange}{\mathlarger{\mathlarger{\mathlarger{\bullet}}}}$  Insula; 
$\textcolor{Cyan}{\mathlarger{\mathlarger{\mathlarger{\bullet}}}}$ Cingulate; 
$\textcolor{LimeGreen}{\mathlarger{\mathlarger{\mathlarger{\bullet}}}}$ Temporal Lobe; $\textcolor{Red}{\mathlarger{\mathlarger{\mathlarger{\bullet}}}}$  Occipital Lobe; $\textcolor{Yellow}{\mathlarger{\mathlarger{\mathlarger{\bullet}}}}$ Parietal Lobe;  $\textcolor{Gray}{\mathlarger{\mathlarger{\mathlarger{\bullet}}}}$ Deep Grey Matter;  $\textcolor{VioletRed}{\mathlarger{\mathlarger{\mathlarger{\bullet}}}}$ Cerebellum.
\label{MSTsnap2}}
\end{figure}

\begin{figure}[!ht]
\vspace{-3mm}
\subfigure[Threshold = 0.50]
{\includegraphics[width=0.5\textwidth]{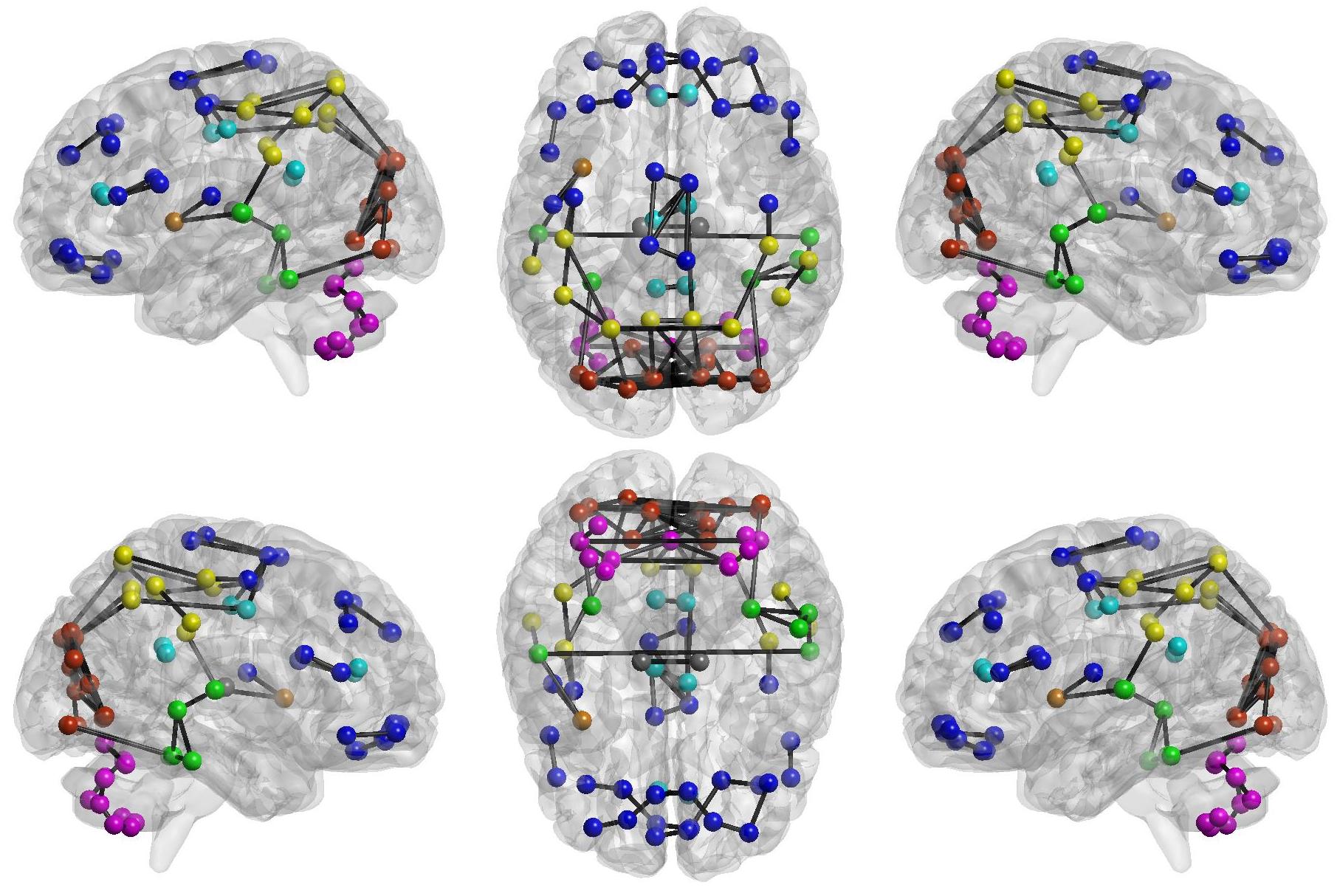}}
\vspace{1mm}
\hspace{1mm}
{\includegraphics[width=0.5\textwidth]{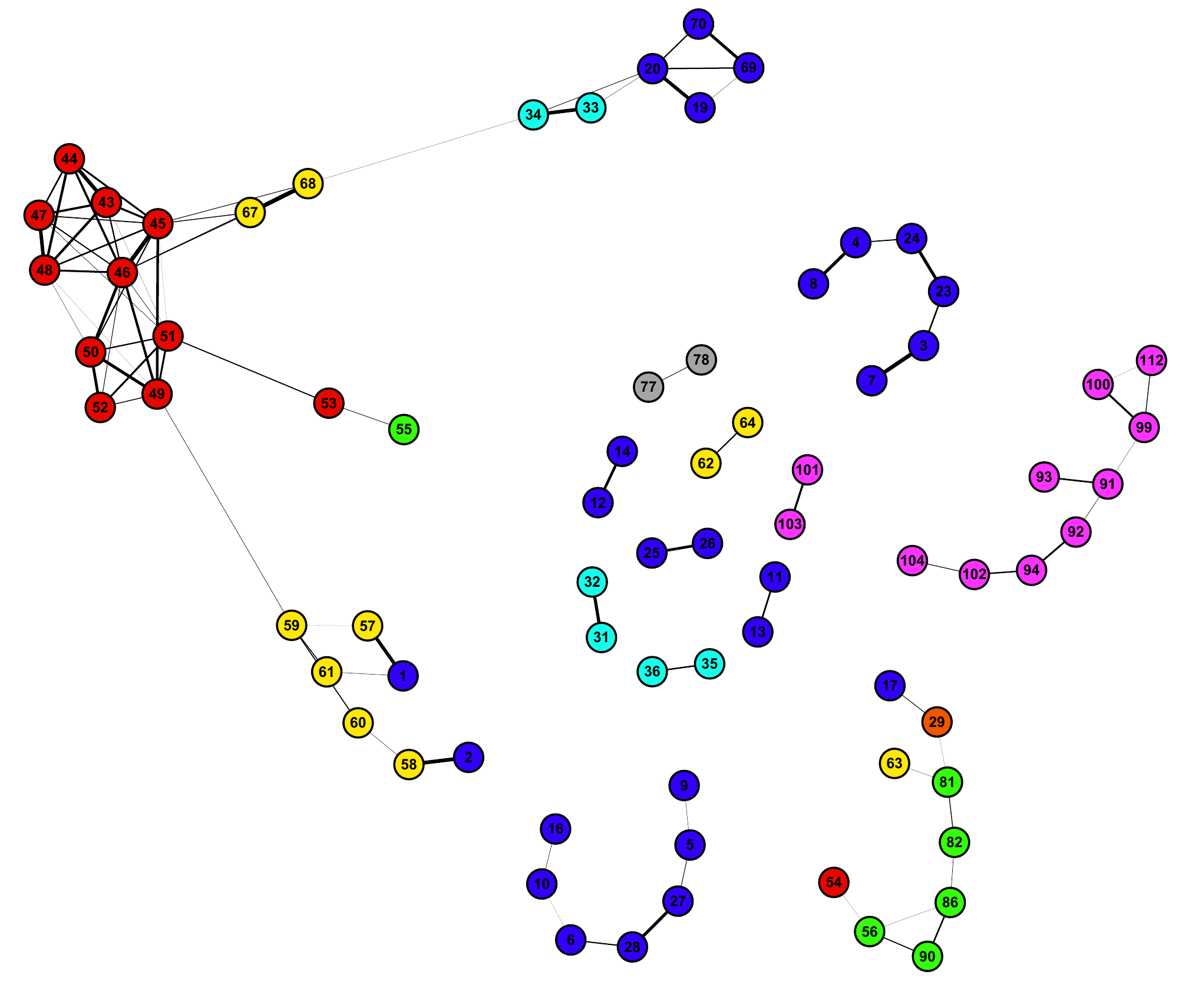}}
\subfigure[Threshold = 0.47]
{\includegraphics[width=0.5\textwidth]{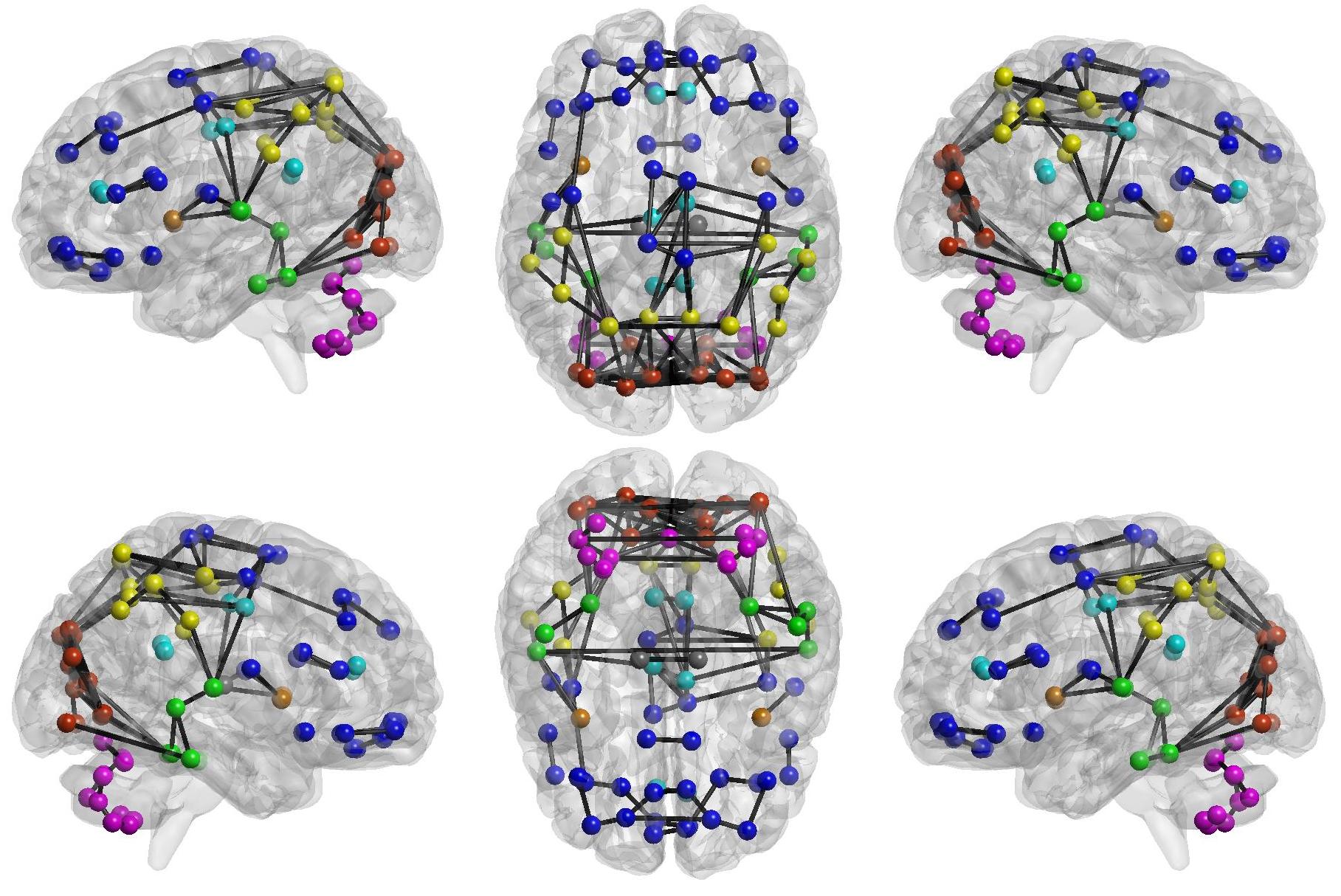}}
\hspace{1mm}
{\includegraphics[width=0.5\textwidth]{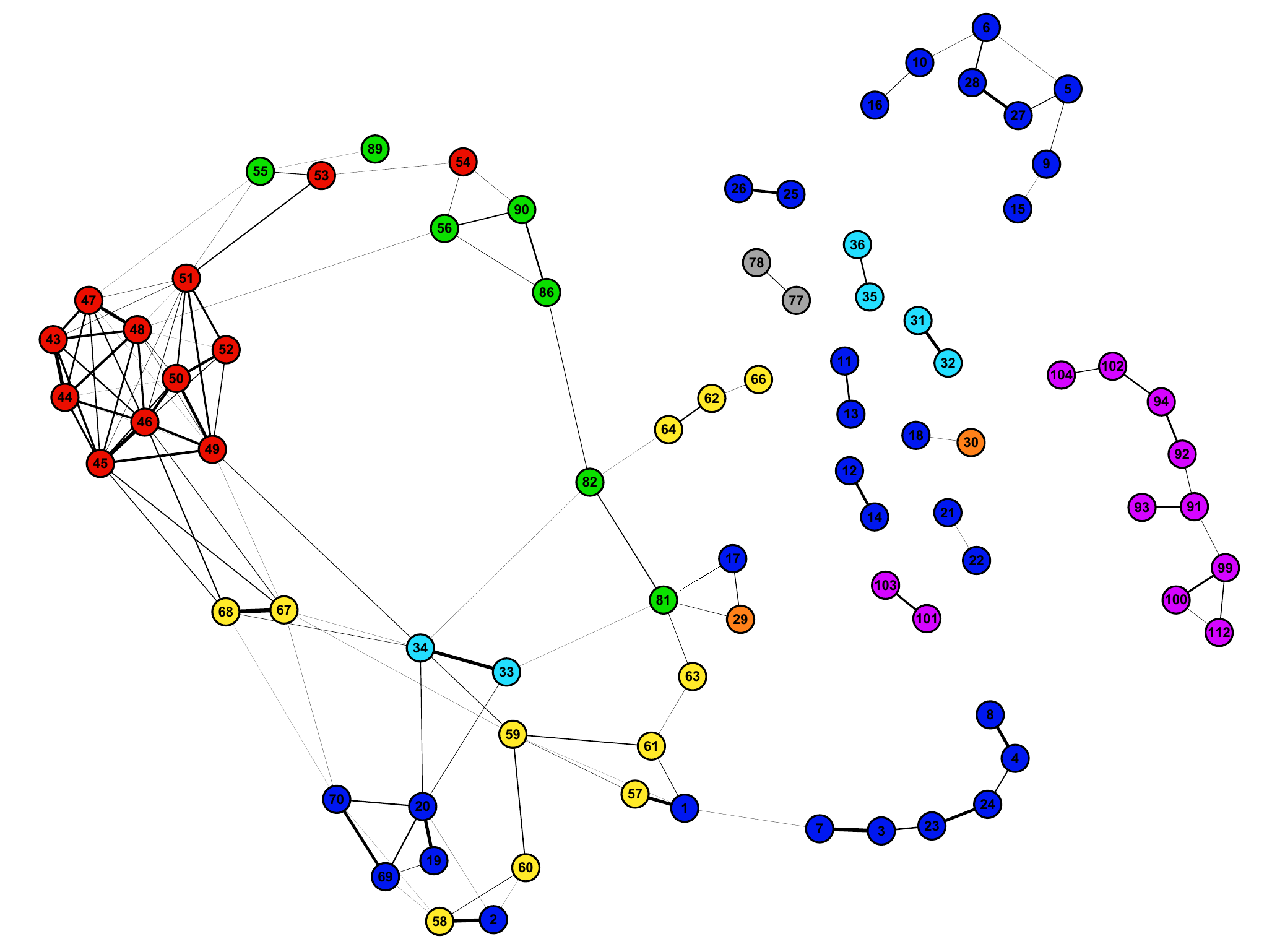}}
\vspace{-3mm}
\subfigure[Threshold = 0.39]
{\includegraphics[width=0.53\textwidth]{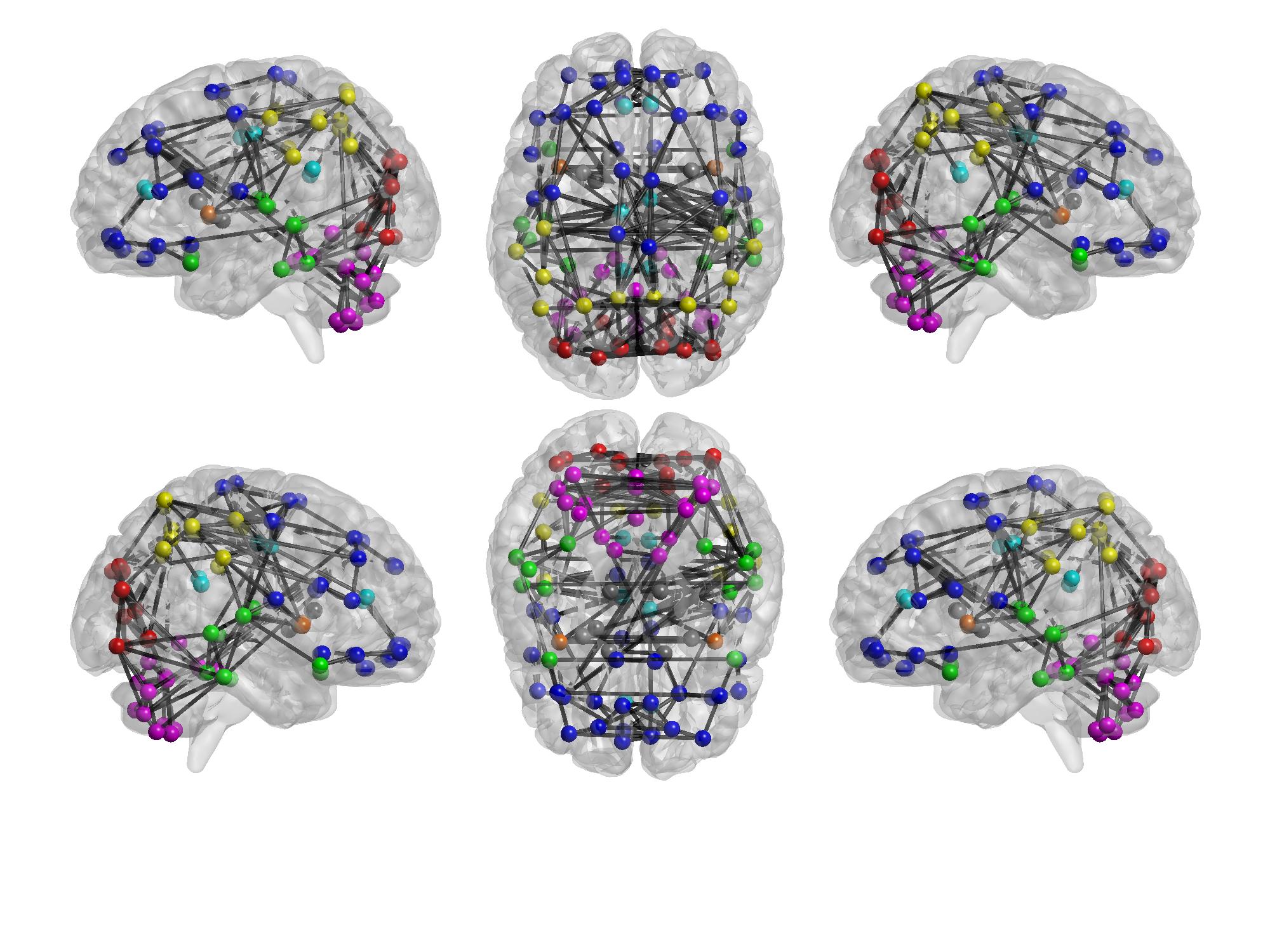}}
\hspace{-4mm}
{\includegraphics[width=0.55\textwidth]{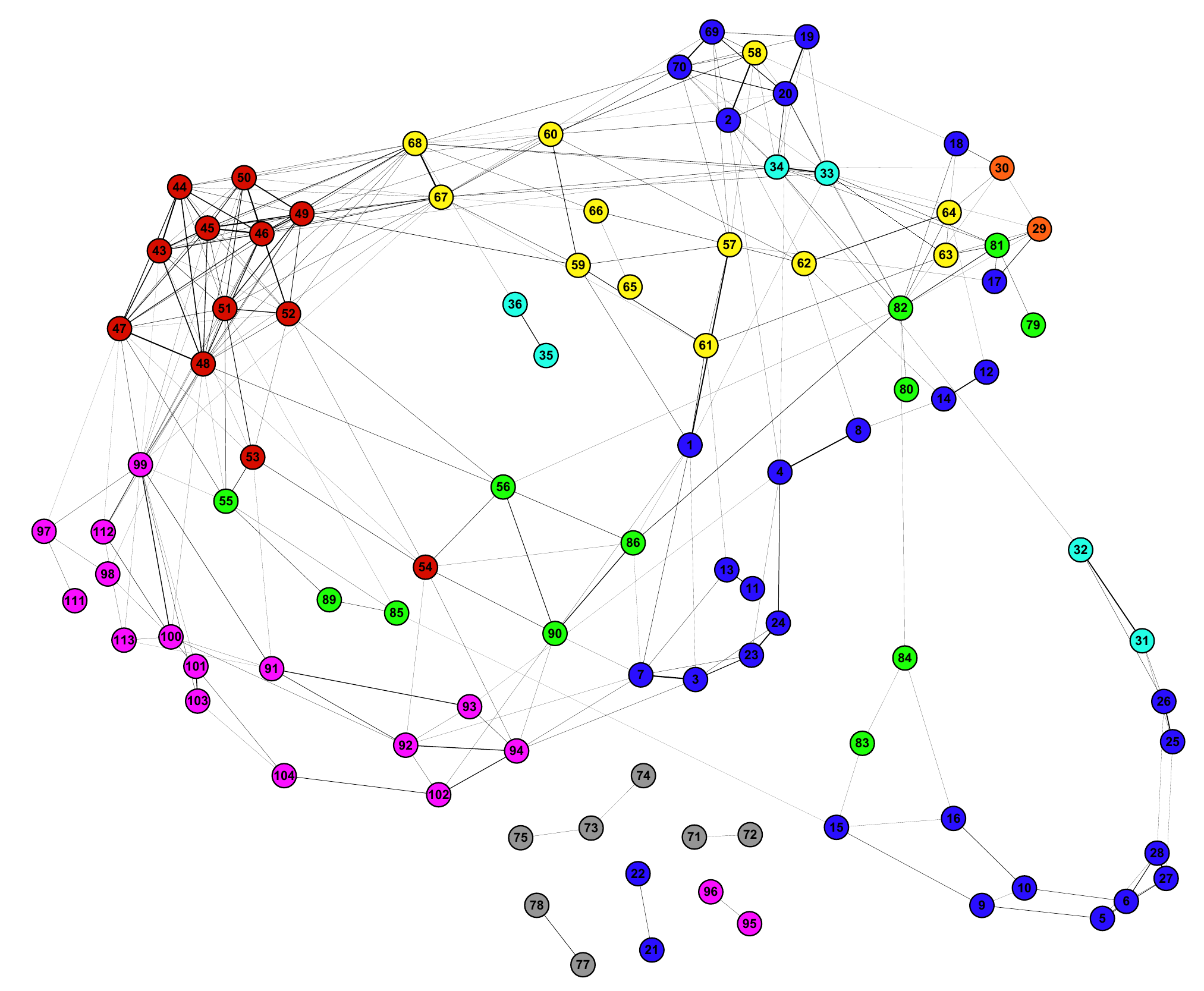}}
\caption{{\bf Snapshots of the human functional brain network.} Each snapshot is realized thresholding the correlation matrix. Thresholds are chosen looking at the most significant deviation of plateaux length of the real percolation curve from its ensemble of randomization (fig.\ref{perc}(b). ) Colors represent anatomical regions according to the grouping of AAL parcellation in fig.\ref{MSF}(a): $\textcolor{Blue}{\mathlarger{\mathlarger{\mathlarger{\bullet}}}}$  Frontal Lobe; $\textcolor{Orange}{\mathlarger{\mathlarger{\mathlarger{\bullet}}}}$  Insula; 
$\textcolor{Cyan}{\mathlarger{\mathlarger{\mathlarger{\bullet}}}}$ Cingulate; 
$\textcolor{LimeGreen}{\mathlarger{\mathlarger{\mathlarger{\bullet}}}}$ Temporal Lobe; $\textcolor{Red}{\mathlarger{\mathlarger{\mathlarger{\bullet}}}}$  Occipital Lobe; $\textcolor{Yellow}{\mathlarger{\mathlarger{\mathlarger{\bullet}}}}$ Parietal Lobe;  $\textcolor{Gray}{\mathlarger{\mathlarger{\mathlarger{\bullet}}}}$ Deep Grey Matter;  $\textcolor{VioletRed}{\mathlarger{\mathlarger{\mathlarger{\bullet}}}}$ Cerebellum.
\label{MSTsnap3}}
\end{figure}



\clearpage
\newpage
\section*{Inter-subject variability}

\begin{figure}[!ht]
\centering
\subfigure[]
{\includegraphics[width=0.35\textwidth]{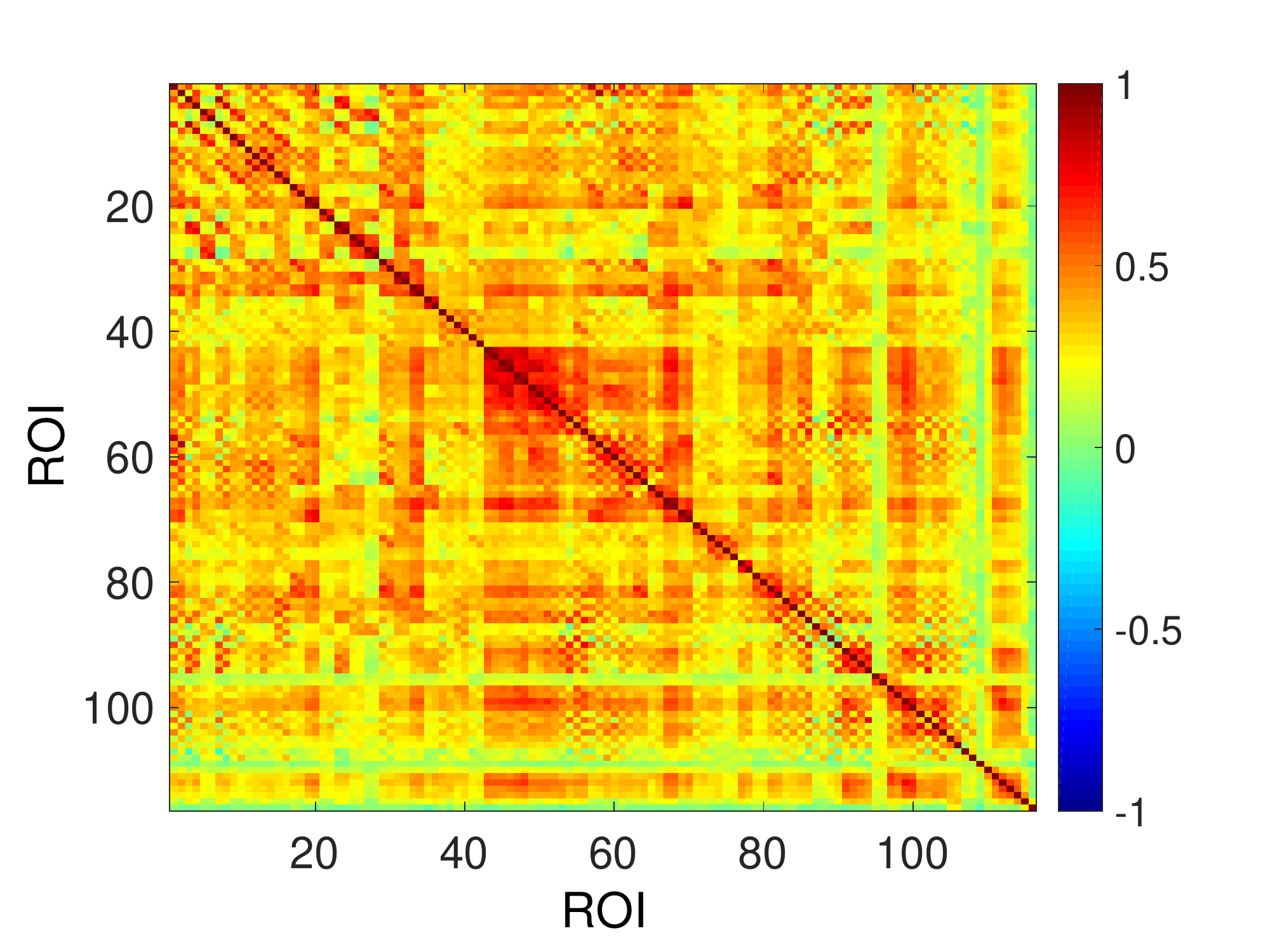}}
\hspace{-7mm}
\subfigure[]
{\includegraphics[width=0.35\textwidth]{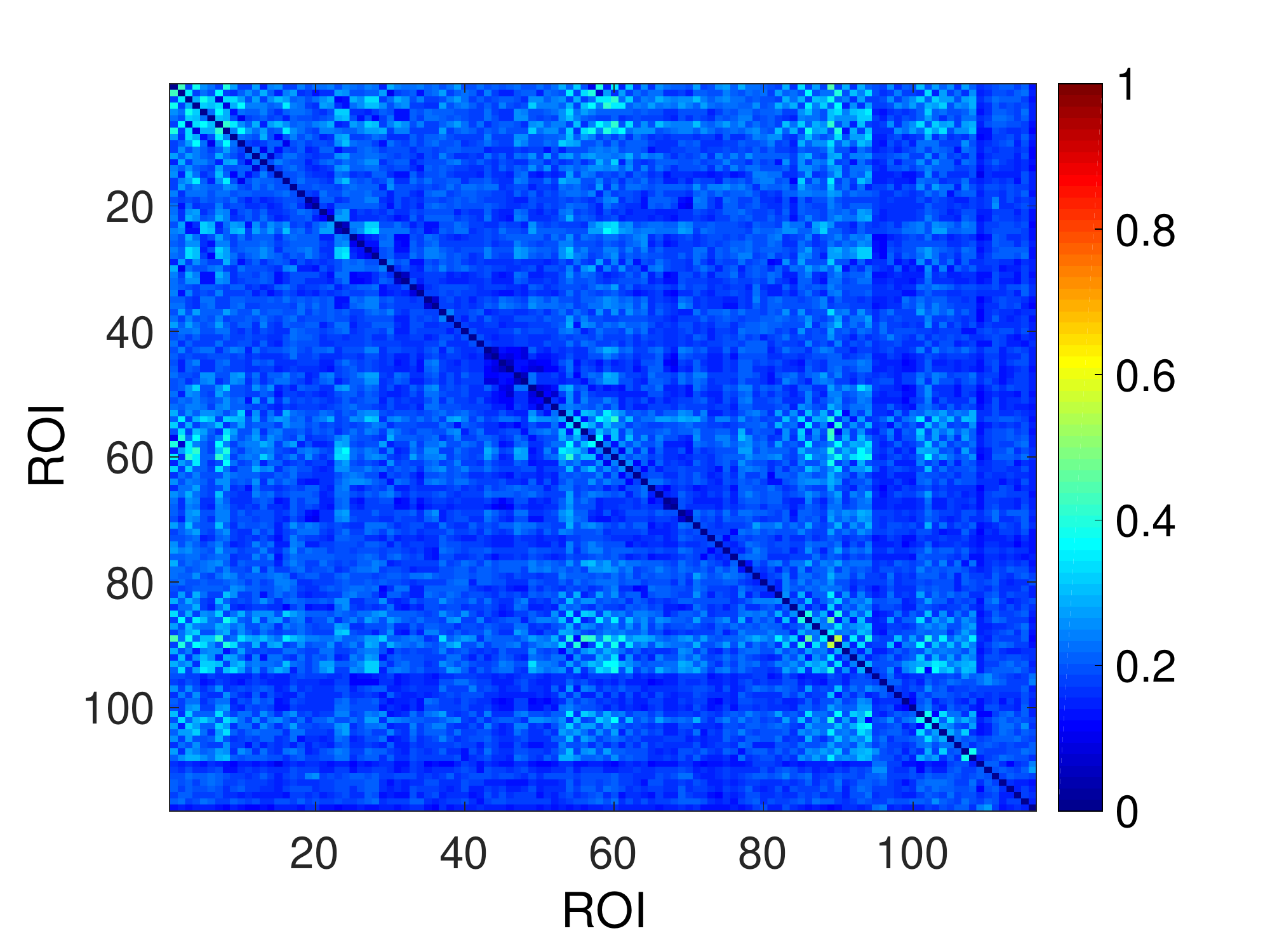}}
\hspace{-7mm}
\subfigure[]
{\includegraphics[width=0.35\textwidth]{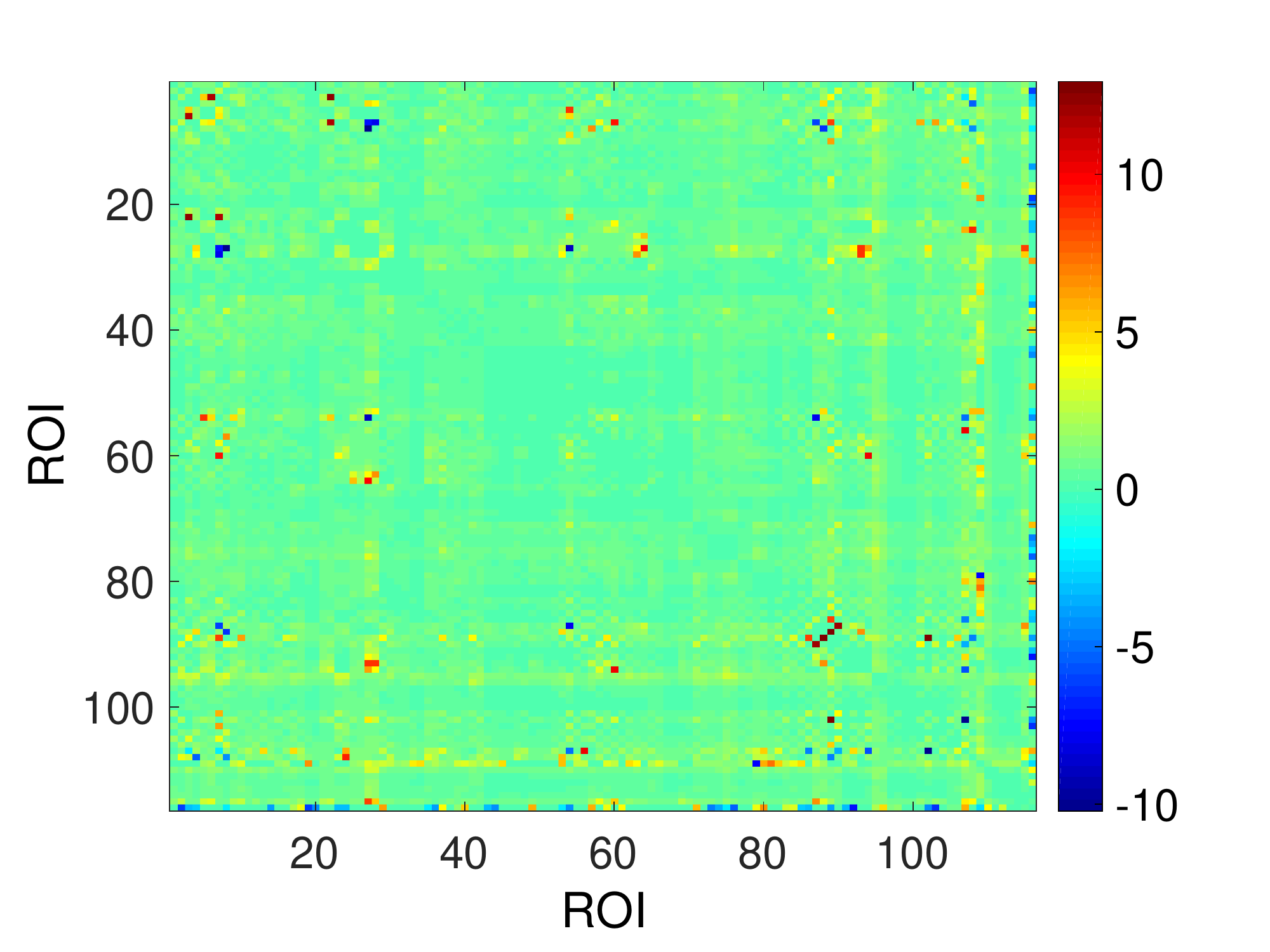}}
\caption{{\bf Inter-subjects variability.} (a) Mean, (b) standard deviation and (c) coefficient of variation of pairwise correlation values across the 40 subjects involved in the study.
\label{std}}
\end{figure}

 We explore the inter-subject variability with several approaches. We compute the standard deviation of pairwise correlation values across the 40 individuals and report it together with the average correlation matrix used for the analysis. Figure \ref{std} (a) shows that there are  few negative correlation values, however small in absolute term. The standard deviation ranges in the  interval $[0.039, 0.55]$, with most of the highest values concerning  correlation averages close to zero (Fig. \ref{std} (b)). Those are few links which show positive and negative values across the 40 individuals and average around zero. They mainly belong to the Cerebellum (see SI) and the inter-subjects variability in this case could be due to the specific anatomical location with consequences on the resulting fMRI. However,  they do not affect our results as in our analysis only the greatest correlation values  play a key-role (percolation, MSF, MST), while the smallest threshold considered in the hierarchical integration of brain areas is 0.33. In figure \ref{std} (c) we also report the coefficient of variation for the correlation matrix thresholded using the False Discovery Rate (FDR). Most of the values concentrate around zero, while the greatest ones correspond to correlation values close to zero.

As next step, we compute the pairwise cosine similarity between the correlation matrices of subjects. We find high similarity between subjects (fig.S15 (a)), with few exceptions (mainly, subjects 35 and 38). However, their presence does not affect the average correlation matrix as explicitly shown by applying the Jackknife resampling technique. One at a time, we remove the correlation matrix corresponding to a subject from the sample and average the remaining 39 correlation matrices. Then, we compute the cosine similarity between the 40 resulting correlation matrices (fig.S15 (b)). The similarity values belong to $[0.9996, 0.9999]$, revealing that the emerged diversity for some subjects is negligible when we compute the average correlation matrix.

\section*{Discussion}

In this paper we have reported the results of a network theory-based investigation of a large human fMRI dataset, aimed at assessing the basal architecture of the resting state functional connectivity network. 
We have computed the Maximal Spanning Forest (MSF) and the Maximal Spanning Tree (MST) for our population-wise brain. The former showed a peculiar composition of the basic modules composing the network, by revealing their structure as linear chains.  The MSF reveals both intra- and inter-hemispherical modules, and the presence of small modules alongside with larger sub-networks. The MST, on the other hand, enables the classification of connectors and provincial areas. 
We also used a modified percolation analysis that retains the information on all the connected components of a given network. Such variation of the percolation analysis does not require the application of a threshold to obtain binary connectivity networks. By means of this technique, applicable to experimental correlation matrices, we showed hierarchically the progressive integration of brain regions in the whole organization of the connectivity network, which does not appear in a null model defined by constraining the spectrum of the observed correlation matrix.
The percolation analysis here used represents a generalization of the classical one \citep{gallos2012small} and is characterised by some improvements \citep{tiz2016} among which it must be recognized the ability to detect newly disconnected modules from secondary components.
As a complex system, the human brain is intrinsically organized into modules, creating a high degree of flexibility and adaptability of the system without fundamentally altering underlying structure \citep{bassett2011understanding}. Remarkably, our percolation analysis unearthed the existence of a hierarchical organization of areas, clusters of areas and modules within the human functional brain network. The process of network aggregation from isolated components appeared slow - compared to the null-model - with the emergence of significant stable configurations identified by several plateaux in the percolation curve. This outcomeda suggests a progressive engagement of brain regions according to the local amount of connectivity, namely a network organization in hierarchical modules whose participation to the entire structure is progressively included as the links strength decreases. The sequence of stable network configurations that, from the single areas, leads to the fully connected network, can be achieved by choosing values associated with specific plateaux of the percolation curve. Although this is technically topological modularity it is remarkable that it reflects functional and anatomical features. For instance bilateral correspondence of regions belonging to the same module was obtained without the introduction of symmetry constraints imposed in the analyses, supporting the conclusion that a hallmark of large-scale resting-state networks is their inter-hemispheric symmetry \citep{smith2009correspondence}. Interestingly, descending the hierarchical staircase of the percolation we found that occipital regions as the cuneus, the lingual gyrus, the occipital superior and middle gyrus and the calcarine sulcus tended to form a densely interconnected module. Along with that, frontal, cerebellar and temporal regions resulted involved in different modules characterised by linear chains, which joined the occipital cluster, in a progressive integration process, through a nodal bridge represented by the precuneus. In particular, temporal and occipital lobes resulted connected thanks to the bridging action of the fusiform and the inferior occipital gyrus as expected by considering the local cytoarchitecture \citep{caspers2013cytoarchitectonical}. Thalamus, putamen, pallidus, hippocampus, amygdala and caudate (here refereed to as deep grey matter) was found to form binary bilateral modules that joined the whole structure at the end of the integration process.  If on one hand the remarkable connectivity within the occipital lobe can be ascribed to an efficient coupling via the interplay of tangential intracortical and callosal connections \citep{gencc2015functional}, on the other hand such connectedness could be ascribed to fixational eye movements \citep{martinez2004role}. In support to this observation, it must be noticed that in Parkinson’s disease patients, which present with eye movement disturbances that accompany the cardinal motor symptoms, the worse the oculomotor performance is, the more the regional functional connectivity is decreased \citep{gorges2016association}.
A second striking result has been obtained from MSF and MST analysis of the whole brain network. For the first time we report a non-trivial and ordered structure for the basal scaffold of the functional network of the brain.
Keeping track only of the strongest connection for each region, the network appeared organised in components exhibiting a linear chain-like structure. When the same procedure has been applied to a proper randomisation of the real human brain functional network, the outcome strongly differed: hubs and star-like structure emerged. This result was tested for robustness by improving the number of regions included in the grey matter parcellation. We calculated correlation matrices from two further templates composed of 276 and 531 regions of interest, respectively, each obtained as a sub-parcellation of the AAL template. The chain-like organization has been proved to persist as the topological outcome of the MSF and the MST analyses of the human brain functional network, against the randomised counterpart that showed a star-like organization (figures S3-S14 in the Supplementary Information). Firstly, modules composition within the MSF deserves a comment. We found linear clusters of regions grouped according neuroanatomical criteria, except the temporal and insular regions. While it is expected to find close portions of the cortex strongly coupled, it is not obvious that the same amount of coupling encompasses regional bilaterality embedded in the modularity of the whole brain network. The exception represented by the temporal and insular regions can be ascribed to different reasons. The insular lobes showed to be more connected with frontal regions, through the Rolandic opercula, than with each other (each one respecting its own laterality). Concurrently, the fronto-insular cortex, wherein generation and experience of emotion are found \citep{gu2013anterior}, is implicated in the elaborate circuitry associated with awareness \citep{craig2009you} and uniquely (together with the anterior cingulate cortex and the dorsolateral prefrontal cortex) characterized by the presence of Von Economo neurons \citep{fajardo2008economo,allman2011economo}. On the other hand the fragmentation of the temporal lobe can follow the mosaic structure of its functional specialization. In fact it includes auditory sensory and association cortices, part of the posterior language cortex, visual and higher order association cortices, primary and association olfactory cortices and enthorinal cortex \citep{frackowiak2004human}. The grouping of other heterogeneous clusters is due both to spatial closeness (precentral and postcentral gyrus) and to homofunctional specialization (hippocampus and parahippocampal gyrus). 
The MST representation of the functional brain network scaffold was found characterized by a long linear structure, whose skeleton follows the anatomical spatial distribution of lobes (from the frontal to the occipital lobe, passing through the temporal and parietal lobes), small chains linked alongside and a star-like structure (composed of cerebellar regions) at the end of one side. The main connectors along the main chain included: the cuneus/precuneus , the superior temporal gyrus, the superior occipital gyrus, the inferior parietal lobule, the medial orbitofrontal cortex/gyrus rectus and the VI lobule of the cerebellum. Notably, precuneus, inferior parietal cortex and medial orbitofrontal cortex constitute the Default Mode Network \citep{greicius2003functional,raichle2007default}, which is a crucial module of the whole brain network both for its functioning and as the first target of metabolic alterations both in neurological and psychiatric diseases \citep{buckner2008brain,broyd2009default}. The large connectivity with the Default Mode Network showed by the superior temporal gyrus, which confers to this region a central role in the MST, has been observed in fMRI data \citep{zhang2012functional} but never found in FDG-PET data. Since one of the differences between the two techniques is the level of noise produced during the recordings (relatively silent during FDG-PET vs intensively noisy during MRI), the absence of overlap of the patterns of connectivity reported in that region might be explained by the auditory engagement included in the MRI experiments \citep{passow2015default}. The same argument works as far as the central role of the superior occipital cortex, being the visual activity engaged by the fixation during the resting-state fMRI experiment. A final comment must be reserved to the cerebellar module. The understanding of human cerebellar function has undergone a paradigm shift. No longer considered purely devoted to motor control, a wide role for the cerebellum in cognitive and affective functions is supported by anatomical, clinical and functional neuroimaging data \citep{picerni2013cerebellum,laricchiuta2014cerebellum}. Recent evidence from functional connectivity studies in humans indicates that the cerebellum participates in functional networks with sensorimotor areas engaged in motor control and with association cortices that are involved in cognitive processes \citep{habas2009distinct,krienen2009segregated,o2010distinct}. The evidence of the cerebellar engagement in higher cognitive functions as well as in sensorimotor processes might explain its central role in the basal scaffold of the whole functional brain network. Specifically, it has been reported that the lobule VI of the cerebellum, here found as a main connector of the network, has been found to be involved in different cognitive tasks, from mental rotation to working memory processing \citep{stoodley2012functional}. In addition, it has been showed that during the first stages of consciousness loss, induced by mild sedation, the pattern of thalamic functional disconnections included regions belonging to the cerebellum \citep{gili2013thalamus}, supporting the evidence of a strong cerebello-thalamic interaction showed by the MST analysis.
Lastly it must be noticed that both in MSF and MST analysis, as weights of our graphs, the square of the correlation coefficient (r2) was used. Positive and negative correlations, characterized by the same absolute value, refer to the same amount of information in terms of connectivity strength. Since in this study we considered undirected weighted graphs, the phase-shift between time series, which gives rise to the sign of the correlation coefficient, was not significantly informative. Indeed, \citet{goelman2014maximizing} showed that both positive and negative correlations are related to synchronized neural activity and that the sign of correlations can result from neural-mediated hemodynamic mechanisms, which lead to temporal and spatial heterogeneity. Moreover, r2 was chosen since it has a strict relationship with the coefficient of determination R2. Specifically, after a simple linear regression, R2 equals the square of the correlation coefficient between the observed and modelled data values \citep{draper1998applied}. In terms of a correlation analysis, r2 estimates how well a time series can predict another one and can be considered as a measure of the magnitude of the relationship between time series.
In conclusion we have reported that the whole architecture of the brain activity at rest is characterised by chain-like structures organized in a hierarchical modular arrangement. This kind of organization is thought to be stable under large-scale reconnection of substructure \citep{robinson2009dynamical}, and efficient in terms of wiring costs \citep{bullmore2012economy}.  In addition the chain-like feature of the basal modules facilitate the wiring and reconnection processes, not asking for a specific region as a hub, but functioning the whole module as a multiple access point for the functional plasticity.

\newpage

\subsection*{Methods}
Forty healthy subjects (mean age=38, SD=10; mean educational attainment=15, SD=3; male N=19) participated in this study. All subjects were carefully screened for a current or past diagnosis of any DSM-5 Axis I or II disorder using the SCID-5 Research Version edition (SCID-5-RV: \citealp{first2015structured}) and the SCID-5 Personality Disorders (SCID-5-PD:\citealp{first2016structured}). Exclusion criteria included: (1) a history of psychoactive substance dependence or abuse during the last one year period evaluated by the structured interview SCID-5-RV, (2) a history of neurologic illness or brain injury, (3) major medical illnesses, that is, diabetes not stabilized, obstructive pulmonary disease or asthma, hematological/oncological disorders, B12 or folate deficiency as evidenced by blood concentrations below the lower normal limit, pernicious anemia, clinically significant and unstable active gastrointestinal, renal, hepatic, endocrine or cardiovascular system disease, newly treated hypothyroidism, (4) the presence of any brain abnormality and microvascular lesion apparent on conventional FLAIR-scans. The presence, severity and location of vascular lesions was computed using a semi-automated method\citep{iorio2013white}, (5) IQ below the normal range according to TIB (Test Intelligenza Breve, Italian analog of the National Adult Reading Test – NART –) (6) global cognitive deterioration according to a Mini-Mental State Examination (MMSE) \citep{folstein1975mini} score lower than 26, (7) dementia diagnosis according with DSM-5 criteria \citep{american2013diagnostic} and (8) non-Italian language native speaker. All participants were right-handed, with normal or corrected-to-normal vision. They gave written informed consent to participate after the procedures had been fully explained. The study was approved and carried out in accordance with the guidelines of the IRCCS Santa Lucia Foundation Ethics Committee.

\subsection*{Data acquisition and preprocessing}
FMRI data were collected using gradient-echo echo-planar imaging at 3T (Philips Achieva) using a (T2*)-weighted imaging sequence sensitive to blood oxygen level-dependent (BOLD) (TR = 3 s, TE = 30 ms, matrix = 80 x 80, FOV=224x224, slice thickness = 3 mm, flip angle = 90°, 50 slices, 240 vol). A thirty-two channel receive-only head coil was used. A high-resolution T1-weighted whole-brain structural scan was also acquired (1x1x1 mm voxels). Subjects were instructed to lay in the scanner at rest with eyes open. For the purposes of accounting for physiological variance in the time-series data, cardiac and respiratory cycles were recorded using the scanner’s built-in photoplethysmograph and a pneumatic chest belt, respectively.
The human brain was segmented into 116 macro-regions from the AAL template \citep{tzourio2002automated}. Resting state fMRI signals were averaged across each region to generate 116 time-series, which in turn were pairwise correlated calculating the Pearson's coefficient and organized in a symmetric matrix. Two further templates were used to test the robustness of results at increasing spatial resolution by subdividing AAL regions as described in \citet{fornito2010network}. The number of regions included in the two templates was 276 and 531 respectively.
Several sources of physiological variance were removed from each individual subject’s time-series fMRI data. For each subject, physiological noise correction consisted of removal of time-locked cardiac and respiratory artifacts (two cardiac harmonics and two respiratory harmonics plus four interaction terms) using linear regression \citep{glover2000image} and of low-frequency respiratory and heart rate effects \citep{birn2006separating,shmueli2007low,chang2009effects}. FMRI data were then preprocessed as follows: correction for head motion and slice-timing and removal of non-brain voxels (performed using FSL: FMRIB’s Software Library, www.fmrib.ox.ac.uk/fsl). Head motion estimation parameters were used to derive the frame-wise displacement (FD): time points with high FD (FD $>$ 0.2 mm) were replaced through a least-squares spectral decomposition as described in \citet{power2014methods}.  Data were then demeaned, detrended and band-pass filtered in the frequency range 0.01-0.1 Hz, using custom software written in Matlab (The Math Works). For group analysis, a two-step registration process was performed. fMRI data were transformed first from functional space to individual subjects’ structural space using FLIRT (FMRIB’s Linear Registration Tool) and then non-linearly to a standard space (Montreal Neurological Institute MNI152 standard map) using Advanced Normalization Tools (ANTs; Penn Image Computing \& Science Lab, http://www.picsl.upenn.edu/ANTS/). Finally data were spatially smoothed (5x5x5 mm full-width half-maximum Gaussian kernel).
In order to create an average adjacency matrix at the population level, subject-wise matrices were Fisher-transformed, averaged across subjects and back-transformed. 
\subsection*{Percolation Analysis and Thresholding}
We used a variation of the percolation analysis proposed by \citet{gallos2012small}, recently developed by \citet{tiz2016}. We ranked all experimentally determined correlation coefficients in increasing order and one at a time we removed from the network the link corresponding to the observed value in order to explore the global organization of the remaining network. by considering the connected components. This enabled us to calculate connected components by systematically varying the threshold on the network.
Among the possible thresholds we looked at those values associated with the plateaux showed in the curve \lq\lq number of connected components vs thresholds\rq\rq{} obtained by the percolation process. Specifically, we considered all correlation values corresponding to points in the distribution of the real plateaux length overcoming the mean plus four standard deviations of the distribution of plateaux length of the ensemble of randomizations. This procedure led to the definition of 23 values, for which stable configurations of brain networks were found. 
\subsection*{Maximum Spanning Forest and Maximum Spanning Tree}
	Starting from the initial correlation matrix we built a new matrix keeping only the maximum values along each row and sending all the rest to zero. We construct a directed network whose links connect each source-ROI with its maximally correlated target-ROI. This approach is equivalent to create a network from scratch in three steps: (i) to rank in decreasing order all the pairwise correlation values; (ii) to look at the list from the top and adding a new link if at least one of the two involved nodes has degree zero, otherwise discarding such correlation; (iii) to update the node degrees. The procedure ends when all nodes have at least degree equal to one. In both cases, the resulting network is made up by several components of nodes connected through their strongest links and without forming loops. In other terms, we have a Maximum Spanning Forest (MSF) of the initial correlation network. Furthermore, the first procedure introduced a directionality, simply revealing for each brain area which is its maximally correlated counterpart. The correlation values discarded during the construction of the MSF and ranked in increasing order were then used to build a Maximum Spanning Tree (MST). Starting from the top of the list, a link between the two involved nodes was drawn if they did not belong to the same group, then the two sets were merged into one; otherwise the correlation value was discarded. The procedure ended when all the nodes belonged to the same connected component.	
\subsection*{A statistical benchmark for human brains}
In order to define to what extent the stepwise structure highlighted by the percolation analysis, the MSF and MSF were significant, we compared the results with a proper statistical benchmark. Specifically, in order to understand whether the results were a proper representation of the real intrinsic organization of the brain functioning at rest we needed to define an appropriate null model.
A matrix has to satisfy some requirements in order to be a correlation matrix: it must be symmetric, with diagonal elements equal to 1, with off-diagonal elements in the range [-1,1] and it has to be positive semidefinite. Not all standard randomization techniques are suitable in this case. Here, we kept fix the spectrum distribution of the observed correlation matrix applying a series of givens rotations. This choice avoids to introduce further procedures to adjust reshuffled matrices in order to be positive semidefinite.  We generated 100 randomizations of the observed correlation matrix. All comparisons between the ensemble of randomizations and the real network were performed on the pointwise squared matrix in order to neglect signs but to keep the ranking of correlation values. Relevant quantities were computed on each randomization, than averaged over the ensemble.

\bibliography{BrainrefRMTGFP}

\bibliographystyle{apalike}

\end{document}